\documentclass{aa}  
\usepackage{natbib}
\usepackage{graphicx}
\usepackage{txfonts}
\usepackage{longtable}
\newcommand{\Msun}{\hbox{$\hbox{M}_\odot\;$}}

\newcommand{\kms}{\hbox{${\rm km}\:{\rm s}^{-1}\;$}}

\newcommand{\kmso}{\hbox{${\rm km}\:{\rm s}^{-1}$}}
\newcommand{\nod}{--}
\newcommand{\teff}{$T_{\rm eff}\;$}  
\newcommand{\teffo}{$T_{\rm eff}$}  
\newcommand{\logg}{$\log\;g\;$}  
\newcommand{\loggo}{$\log\;g$}

\newcommand{\cobold}{\ensuremath{\mathrm{CO}^5\mathrm{BOLD}}}
\newcommand{\tact}{($\langle$3D$\rangle$ -- 1D$_\mathrm{MARCS}$)}
\newcommand{\ftac}{3D -- 1D$_{\rm LHD}$}
\newcommand{\tac}{3D -- $\langle$3D$\rangle$}
\newcommand{\linfor}{Linfor3D}

\newcommand{\pun}[1]{\,#1}

\newcommand{\mlp}{\ensuremath{\alpha_{\mathrm{MLT}}}}

\bibpunct{(}{)}{;}{a}{}{,} 

\begin{document}
   \title{First Stars XI. Chemical composition of the extremely metal-poor dwarfs in
the binary CS 22876--032\thanks {Based on observations made with the ESO Very Large Telescope 
at Paranal Observatory, Chile (Large Programme ``First Stars'', 
ID 165.N-0276(A); P.I. R. Cayrel).}}

   \subtitle{}

   \author{J. I. Gonz\'alez Hern\'andez\inst{1,2},  P.
   Bonifacio\inst{1,2,3}, H.-G. Ludwig\inst{1,2}, E. Caffau\inst{2}
   \and M. Spite\inst {2} \and
F. Spite \inst {2} \and
R. Cayrel\inst {2} \and
P. Molaro\inst {2,3} \and
V. Hill\inst {2} \and 
P. Fran\c cois  \inst{2} \and
B. Plez\inst {5} \and
T.~C.~Beers\inst {4} \and
T. Sivarani \inst{4} \and
J.~Andersen\inst {6,7} \and
B.~Barbuy\inst {8} \and
E. Depagne \inst{9} \and
B.~Nordstr\"om \inst {6} \and 
F. Primas\inst {10}
} 

   \offprints{J. I. Gonz\'alez Hern\'andez}

   \institute{
   CIFIST Marie Curie Excellence Team 
         \and
GEPI, Observatoire de Paris, CNRS, Universit\'e Paris Diderot; Place
Jules Janssen 92190
Meudon, France \\
\email{[Jonay.Gonzalez-Hernandez;Piercarlo.Bonifacio;Hans.Ludwig;Elisabetta.Caffau;Roger.Cayrel;\\
;Monique.Spite;Francois.Spite;Vanessa.Hill;Patrick.Francois]@obspm.fr}
         \and
   Istituto Nazionale di Astrofisica - Osservatorio Astronomico di
   Trieste, Via Tiepolo 11, I-34143  Trieste, Italy
  \email {molaro@ts.astro.it}
\and
 Department of Physics and Astronomy, CSCE: Center for the Study of 
Cosmic Evolution, and JINA: Joint Institute for Nuclear Astrophysics, 
Michigan State University, E. Lansing, MI  48824, USA\\
   \email {[thirupati;beers]@pa.msu.edu}
         \and
GRAAL, Universit\'e de Montpellier II, F-34095 
Montpellier
             Cedex 05, France\\
   \email {Bertrand.Plez@graal.univ-montp2.fr}
         \and
         The Niels Bohr Institute, Astronomy, Juliane Maries Vej 30,
         DK-2100 Copenhagen, Denmark\\
   \email {[ja;birgitta]@astro.ku.dk}
\and
     Nordic Optical Telescope, Apartado 474, E-38700 Santa Cruz de 
     La Palma, Spain\\
   \email {ja@not.iac.es}
         \and
    Universidade de Sao Paulo, Departamento de Astronomia,
Rua do Matao 1226, 05508-900 Sao Paulo, Brazil\\
   \email {barbuy@astro.iag.usp.br}
               \and
Las Cumbres Observatory, Goleta, CA 93117, USA\\
\email{edepagne@lcogt.net}
        \and 
         European Southern Observatory (ESO),
         Karl-Schwarschild-Str. 2, D-85749 Garching b. M\"unchen, 
Germany\\
   \email {fprimas@eso.org}
}

   \titlerunning{Chemical abundances of EMP dwarfs in CS 22876--032}
   \authorrunning{Gonz\'alez Hern\'andez et al.}

   \date{Received July, 2007; accepted --}

 
\abstract
{Unevolved metal-poor stars constitute a fossil record of the Early 
Galaxy, and can provide invaluable information on the properties of the
first generations of stars. Binary systems also provide direct
information on the stellar masses of their member stars.}  
{The purpose of this investigation is a detailed abundance study of the
double-lined spectroscopic binary CS 22876--032, which comprises the
two most metal-poor dwarfs known.
} 
{We have used high-resolution, high-S/N ratio spectra from the
UVES spectrograph at the ESO VLT telescope. Long-term radial-velocity 
measurements and broad-band photometry 
allow us to determine improved orbital elements and stellar
parameters for both components. We use OSMARCS 1D models and the
{{\scshape turbospectrum}} spectral synthesis code to determine the 
abundances of Li, O, Na, Mg, Al, Si, Ca, Sc, Ti, Cr, Mn, Fe, Co and
Ni. We also use the CO$^5$BOLD model atmosphere code to compute the 3D
abundance corrections, especially for Li and O.} 
{We find a metallicity of [Fe/H]$\sim -3.6$ for both stars, using 1D
models with 3D corrections of $\sim -0.1$ dex from averaged 3D models.
We determine the oxygen abundance from the near-UV OH bands; the 3D 
corrections are large, $-1$ and $-1.5$ dex for the secondary and
primary respectively, and yield [O/Fe] $\sim 0.8$, 
close to the high-quality results obtained from the [OI] 630 nm 
line in metal-poor giants. Other [$\alpha$/Fe] 
ratios are consistent with those measured in other dwarfs and giants 
with similar [Fe/H], although Ca and Si are somewhat low ([X/Fe]$\la 0$). 
Other element ratios follow those of other halo stars. The Li abundance 
of the primary star is consistent with the Spite plateau, but the secondary 
shows a lower abundance; 3D corrections are small. } 
{The Li abundance in the primary star supports the extension of the
{{\em Spite Plateau}} value at the lowest metallicities, without any
decrease. The low abundance in the secondary star could be explained
by endogenic Li depletion, due to its cooler temperature. If this is
not the case, another, yet unknown mechanism may be causing 
increased scatter in A(Li) at the lowest metallicities. }  


\maketitle
%

\section{Introduction}

Extremely metal-poor (EMP) stars formed with the chemical composition of the gas
in the early Galaxy, and constitute a unique source of information on the first
generations of stars. Among EMP stars, a special place is held by the dwarfs,
which are not subject to the mixing episodes experienced by giants, thus
enhancing their value as cosmological probes. 

In fact, among these stars the Li abundance appears to be constant
whatever the stellar temperature or metallicity \citep{sas82a,sas82b}, the {\em Spite plateau}.
The simplest interpretation of the plateau is that it represents the primordial
Li abundance, i.e., it reflects the amount of Li formed in the first
minutes of the existence of the Universe. If so, Li can be used as a
``baryometer'', a tool to measure the baryonic density of the Universe, since
this is the only cosmological parameter upon which the primordial Li abundance
depends upon.

The independent determination of the baryonic density from the fluctuations of
the Cosmic Microwave Background by the WMAP satellite \citep{spe03,spe07} and
other CMB experiments measuring fluctuations on smaller angular scales, such as
the VSA \citep{reb04, gra03}, ACBAR \citep{kuo04} and CBI
\citep{pea03} experiments, implies a primordial Li abundance which is at least a
factor of 3--4 larger than that observed on the {\em Spite plateau},
creating a conflict with the traditional interpretation of the
plateau. 

In Paper VII in this series \citep{bon07} we have investigated the {\em Spite
plateau} at the lowest metallicities (down to [Fe/H]=--3.3) and found marginal
evidence that at these low metallicities there could be an increased
scatter or even a sharp drop in the Li abundance. It is therefore of 
great interest to explore the Li abundance in stars of even lower
metallicity.  

The star CS 22876--032 was identified in the first paper reporting results of
the HK objective-prism survey by \citet{bee85}, who noted that it had the
weakest Ca II K line in the low-metallicity sample, suggesting that it could be
as metal-deficient as the record holder at that time, the giant CD
-38$^\circ$245 \citep{ban84}. CS 22876--032 had already been observed in the
objective-prism survey of \citet{sle71}, who classified it as an
A-type peculiar star and noted its weak and diffuse Balmer
lines. Having assigned to this star a much earlier spectral type, they
did not conclude that the weakness of the Ca II K line was indeed due
to an extremely low metallicity.   

At the conference ``Chemical and Dynamical Evolution of Galaxies'' in 1989
\citep{bon89}, P. Molaro announced that high-resolution spectra from the CASPEC
spectrograph at the ESO 3.6\pun{m} telescope indicated [Fe/H] $\sim -4.3$ for CS
22876--032. However, just afterwards \citet{nis89} discovered, from simlar-
resolution spectra, that the star is a double-lined spectroscopic binary. The
spectra acquired by Molaro were obtained at a single-lined phase, and the
abundance analysis of CS 22876--032 by \citet{mac90} assumed that it was a
single star. Thus, veiling was neglected, the adopted temperature was too low,
and the measured [Fe/H] was therefore a lower limit to the metallicity of the
system.   

Although CS 22876--032 is relatively bright (V=12.84) for an EMP star,
it took another ten years before a sufficient number of high-resolution 
spectra had been accumulated to allow to determine of the orbital parameters
of this system, and to perform a consistent chemical analysis. \citet{nor00}
found the orbital period to be  
424.7 days and the metallicity of the system [Fe/H]=--3.71. In spite of the
upward revision of the metallicity, partly due to the different solar
Fe abundance assumed ($\log \epsilon(\mathrm{X})_{\odot}=7.50$ instead
of 7.63 in \citealt{mac90}), the two stars in CS 22876--032 remain 
the most metal-poor dwarfs known.  

Thus, the CS 22876--032 system constitutes a unique fossil, recording the
chemical composition of the early Galaxy. Moreover, it allows a measurement of
the Li abundance which probes the {\em Spite plateau} at a lower metallicity
than any other known dwarfs. Note that, despite its the extremely low iron
abundance ([Fe/H]=--5.4), the star HE 1327--2326 \citep{fre05} has very high C, N
and O abundances, so its global metallicity, $Z$, is considerably higher than
that of CS 22876--032. It has also been shown recently that this star is most
likely a slightly evolved subgiant, not a dwarf. 
 
In this paper we use high-resolution, high-S/N ratio spectra from the ESO Kueyen
8.2m telescope and the UVES spectrograph to improve the orbital solution and
perform a complete chemical analysis of the two stars that comprise
CS~22876--032. With respect to the \citet{nor00} analysis, our superior S/N ratio and larger spectral coverage
permit measurement of abundances for many more elements, and, most importantly, for
{\em both} components; the \citet{nor00} analysis of the secondary star was
limited to Fe.

\begin{figure}[ht]
\centering
\includegraphics[clip=true,height=9.5cm,angle=90]{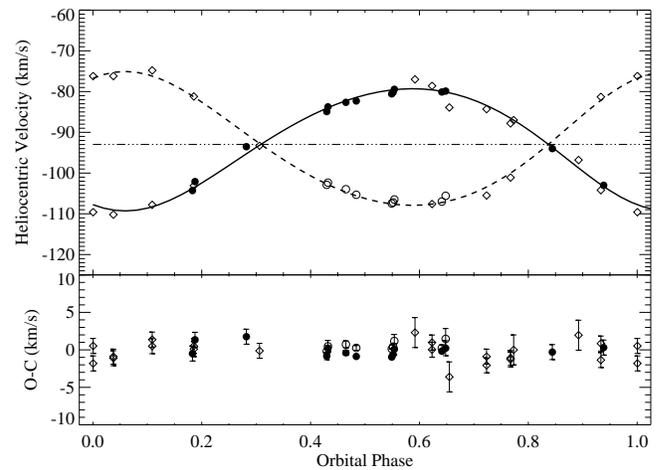}
\caption{ Upper panel: Radial velocities of CS 22876--032 (filled circles: A; open circles: B). Open diamonds: Data from Norris et al. (2000). The curves 
show the orbital solution ($P$ = 425 d, $e$ = 0.14) for Star A (solid) and B 
(dashed). Dot-dashed horizontal line: Centre-of-mass velocity of the system.
Lower panel: Residuals from the fit.} 
\label{vrad}
\end{figure}

\begin{table*}[th]
\caption{Radial-velocity observations of CS 22876--032. For each
velocity, $\sigma$ is the estimated error and {\it (O-C)} the residual
from the orbital fit. $^\star$: $H_\alpha$ velocities; omitted from
solution.}  
\centering                          
\begin{tabular}{lrrrrrrrrr}        
\noalign{\smallskip}
\noalign{\smallskip}
\hline
\hline       
\noalign{\smallskip}
Date & HJD-2,400,000 & Phase & $V_A$ & $\sigma V_A$ & $(O-C)_A$ & $V_B$ & $\sigma V_B$ & $(O-C)_B$ & Source \\    
(days) & & \kmso &  \kmso & \kmso & \kmso & \kmso & \kmso & \\
\noalign{\smallskip}
\hline                        
\noalign{\smallskip}
1985 Sep 6 ....... & 46315.1 & 0.669 & -83.9  & 2.0 &  -3.44  &\nod   & \nod &   \nod & NBR00 \\
1985 Dec 16 ...... & 46416.0 & 0.907 & -96.8  & 2.0 &	1.72  &\nod   & \nod &   \nod & NBR00 \\
1988 Sep 27 ...... & 47431.7 & 0.297 & -93.5  & 1.0 &	1.44  &\nod   & \nod &   \nod & MC90  \\
1989 Sep 13 ...... & 47783.2 & 0.124 & -107.8 & 1.0 &	0.63  &-74.8  & 1.0  &   1.48 & NBR00 \\
1989 Oct 15 ...... & 47814.6 & 0.199 & -104.3 & 1.0 &  -0.94  &\nod   & \nod &   \nod & NBR00 \\
1989 Oct 16 ...... & 47815.7 & 0.201 & -103.1 & 1.0 &	0.07  &-81.2  & 1.0  &   0.86 & NBR00 \\
1989 Oct 17 ...... & 47816.6 & 0.204 & -102.1 & 1.0 &	0.87  &\nod   & \nod &   \nod & NBR00 \\
1989 Dec 6 ....... & 47867.0 & 0.322 & -93.3  & 1.0 &  -0.35  &\nod   & \nod &   \nod & NBR00 \\
1990 Sep 1 ....... & 48135.7 & 0.955 & -103.0 & 1.0 &	0.83  &\nod   & \nod &   \nod & CASPEC \\
1990 Sep 27 ...... & 48162.1 & 0.017 & -109.6 & 1.0 &  -0.91  &-76.2  & 1.0  &  -0.20 & NBR00 \\
1991 May 23 ...... & 48400.0$^\star$ & 0.577 & -87.5  & 2.0 &  -7.34  &\nod   & \nod &   \nod & EMMI \\ 
1991 Aug 22 ...... & 48490.5 & 0.790 & -87.0  & 2.0 &  -0.47  &\nod   & \nod &   \nod & NBR00 \\
1991 Sep 21 ...... & 48520.7 & 0.861 & -94.0  & 1.0 &  -0.65  &\nod   & \nod &   \nod & EMMI \\ 
1992 Aug 17 ...... & 48851.8 & 0.640 & -78.6  & 1.0 &	1.41  &-107.6 & 1.0  &  -0.12 & NBR00 \\
1992 Dec 12 ...... & 48967.5$^\star$ & 0.912 & -90.9  & 2.0 & 8.33 & \nod & \nod & \nod & EMMI \\ 
1996 Aug 7 ....... & 50303.1 & 0.057 & -110.2 & 1.0 &  -0.28  &-76.2  & 1.0  &  -1.55 & NBR00 \\
1997 Aug 23 ...... & 50683.8 & 0.957 & -104.2 & 1.0 &  -0.57  &-81.3  & 1.0  &   0.25 & NBR00 \\
1998 Aug 12 ...... & 51038.3 & 0.787 & -87.8  & 1.0 &  -1.47  &-101.1 & 1.0  &  -0.56 & NBR00 \\
1999 Jul 29 ...... & 51388.7 & 0.612 & -77.0  & 2.0 &	2.89  &\nod   & \nod &   \nod & NBR00 \\ 
1999 Sep 23 ...... & 51444.6 & 0.744 & -84.3  & 1.0 &  -0.93  &-105.5 & 1.0  &  -1.71 & NBR00 \\
2000 Jul 19 ...... & 51744.8 & 0.451 & -84.9  & 0.5 &  -0.52  &-102.9 & 0.3  &  -0.22 & VLT \\
2000 Jul 20 ...... & 51745.8 & 0.453 & -83.7  & 0.4 &	0.56  &-102.3 & 0.8  &   0.51 & VLT \\
2000 Aug 3 ....... & 51759.8 & 0.486 & -82.6  & 0.3 &	0.13  &-103.9 & 0.5  &   0.59 & VLT \\
2000 Aug 11 ...... & 51767.9 & 0.505 & -82.3  & 0.3 &  -0.31  &-105.3 & 0.3  &   0.01 & VLT \\
2000 Oct 17 ...... & 51834.7 & 0.662 & -80.1  & 0.3 &	0.22  &-107.0 & 0.4  &   0.14 & VLT \\
2000 Oct 20 ...... & 51837.6 & 0.669 & -79.9  & 1.0 &	0.56  &-105.6 & 1.4  &   1.38 & VLT \\
2001 Nov 7 ....... & 52220.6 & 0.571 & -80.6  & 0.3 &  -0.35  &-107.4 & 0.3  &  -0.18 & VLT \\
2001 Nov 8 ....... & 52221.6 & 0.573 & -80.2  & 0.3 &	0.01  &-107.2 & 0.3  &   0.06 & VLT \\
2001 Nov 9 ....... & 52222.6 & 0.575 & -79.4  & 0.7 &	0.78  &-106.6 & 0.9  &   0.69 & VLT \\
\noalign{\smallskip}
\hline                              	       
\noalign{\smallskip}
\noalign{\smallskip}
\noalign{\smallskip}
\end{tabular}
\begin{minipage}[r]{17cm}
Sources: NBR00: \citet*[and references therein]{nor00}; MC90: \citet{mac90}; CASPEC: Unpublished velocity from a Caspec spetrum; EMMI: Unpublished velocities from EMMI spectra; VLT: This paper.
\end{minipage}
\normalsize
\rm
\label{tbl1}  
\end{table*}

\section{Observations and data reduction}

Spectroscopic observations of the CS~22876--032 were carried out with the
UV-Visual Echelle Spectrograph (UVES, Dekker et al. 2000) at the
European Southern Observatory (ESO), {\itshape Observatorio Cerro
Paranal}, using the 8.2 m VLT-Kuyen telescope on 2000 July 19, 20,
August 3, 11, and October 17, 20, and 2001 November 7, 8 and 9,
covering the spectral region from $300.0$ nm to $1040.0$ nm. Most
of the observations were made with a projected slit width of
1$\arcsec$ at a resolving power $\lambda/\delta\lambda\sim43\,000$. The
spectra were reduced in a standard manner using the UVES reduction 
package within the {\scshape MIDAS} environment. The signal-to-noise
ratio per pixel varies from 25 at 312.0 nm, 50 at 330.0 nm up to
150 or higher above 410.0 nm.

\begin{table}[!ht]
\centering
\caption[]{Orbital elements of CS 22876--032}
\begin{tabular}{lcc}     
\noalign{\smallskip}
\noalign{\smallskip}
\noalign{\smallskip}
\hline
\hline       
\noalign{\smallskip}
Parameter & This paper & Norris et al. (2000)  \\
\noalign{\smallskip}
\hline
\noalign{\smallskip}
$P$ (days)                 & $424.81 \pm 0.37$ & $424.71 \pm 0.60$ \\ 
$T_0$ (HJD$-$2,400,000)    & $48579.8 \pm 7.1$ & $48576.4 \pm 13.5$\\ 
$e$                        & $0.143 \pm 0.013$ & $0.12 \pm 0.03$   \\ 
$w$ (deg)                  & $148.1 \pm 6.4$   & $144.96 \pm 12.4$ \\ 
$V_0$ (\kmso)              & $-93.11 \pm 0.13$ & $-93.36 \pm 0.28$ \\ 
$K_A$ (\kmso)              & $15.04 \pm 0.26$  & $15.13 \pm 0.51$	  \\ 
$K_B$ (\kmso)              & $16.51 \pm 0.27$  & $17.06 \pm 0.56$	  \\ 
$M_A \sin^3 i (M_{\odot})$ & $0.701 \pm 0.021$ & $0.76 \pm 0.04$   \\
$M_B \sin^3 i (M_{\odot})$ & $0.639 \pm 0.019$ & $0.68 \pm 0.04$   \\
$M_B/M_A$                  & $0.911 \pm 0.022$ & $0.89 \pm 0.04$   \\ 
$\sigma$ (\kmso)           &  1.01             &  1.60             \\
\noalign{\smallskip}
\hline
\hline
\noalign{\smallskip}
\noalign{\smallskip}
\noalign{\smallskip}
\end{tabular}
\normalsize
\rm
\label{tbl2}  
\end{table}    

\section{Revised orbital parameters}

We derived radial velocities from the UVES spectra by fitting a Gaussian
to several unblended spectral lines within the {\scshape IRAF}\footnote{IRAF is
distributed by National Optical Astronomy Observatory, which is operated by the
Association of Universities for Research in Astronomy, Inc., under contract with
the National Science Foundation.} context. Table~\ref{tbl1} contains the radial
velocities and the 1-$\sigma$ errors estimated from the dispersion of
the measurements of different stellar lines. We also list other velocity data
given by \citet*{nor00} and references therein, or which we have measured from
previously unpublished CASPEC or EMMI spectra of this system. 

Here and in the rest of the paper, we denote the more massive and
luminous primary star as Star A, the secondary as Star B. 

Our new radial-velocity measurements of CS 22876--032 extend the time
coverage of its orbit considerably and permit improvement of the
orbital elements relative to those by published by \citet{nor00}. The
computed orbital parameters are given in Table~\ref{tbl2};
Fig.~\ref{vrad} compares the observed radial velocities of both stars
with the curves predicted from these orbital elements.  

Note especially the improved mass ratio, $M_B/M_A=0.91 \pm 0.02$,
which provides stringent constraints on the stellar parameters for the
two components of the binary as discussed below. Note also that the
orbital eccentricity is the lowest found among halo spectroscopic
binaries with periods in the range 100-2000 days (see
\citealt{lat02,gol02}). While this might be a hint that tidal
interaction has been strong in this system, perhaps in the
pre-main-sequence phase, the separation of the stars has been so large
throughout their main-sequence life that this is unlikely to be the
cause of the apparent Li depletion we find in star B (see Sect.
\ref{li_disc}).   

\section{Model Atmospheres}

\subsection{One-dimensional models\label{s1dm}}

Our analysis used {\scshape OSMARCS} 1D LTE model
atmospheres \citep{gus75,ple92,edv93,asp97,gus03} and the {\scshape 
turbospectrum} spectral synthesis code \citep{aap98}. Models were
interpolated in pre-computed grids for a metallicity of $[{\rm
Fe}/{\rm H}]=-3.6$, since this was the final iron abundance (see
\S~\ref{sabu}), and with an $\alpha$-element enhancement of
$[\alpha/\mathrm{Fe}]= +0.4$ dex. We adopted solar abundances from 
\citet{gas00}, with the exception of O, 
for which we adopted
$\log \epsilon(\mathrm{O})_{\odot}=8.72$, 
based on 3D  model atmospheres
\citep{LS07,caffau07}.
The code {\scshape turbospectrum} is used to determine 1D element
abundances in each component of the binary, either via spectrum
synthesis or by comparing the observed equivalent widths of different
stellar lines with the theoretical curves of growth (see
\S~\ref{sabu}).    

\begin{figure}[ht]
\centering
\includegraphics[clip=true,width=9cm]{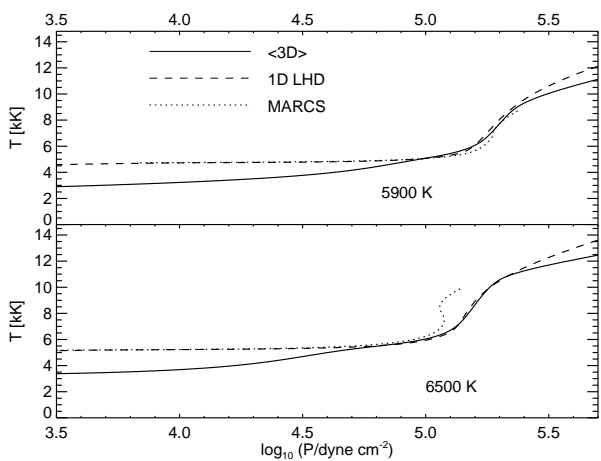}
\caption{Average temperature profile of 3D \cobold\ atmospheric models
(solid lines) compared to 1D$_{\rm LHD}$ (dashed) and 1D MARCS models 
(dotted) in stars A (bottom panel) and B (top).}
\label{mod3d}
\end{figure}

\subsection{Three-dimensional models\label{s3dm}}

In this work we also consider two 3D model atmospheres, which have been
computed with the \cobold\ code \citep{fsd02,wed03}, one for each star. The
atmospheric parameters are close to those observed for the two stars:
\teff/\logg/[Fe/H]: 6550/4.50/--3.0 (A) and 5920/4.50/--3.0 (B). Each model
consists of a representative set of snapshots sampling the temporal evolution
of the photospheric flow at equal intervals in time. The total time intervals
were 2400\pun{s} for the warmer star A, and 9500\pun{s} for the cooler star B.
These time intervals should be compared to the convective turn-over
timescales. From the hydrodynamical point of view, typical timescales in
the models for both components are not much different from that in a solar
model, where the convective turn-over timescale amounts to about 500\pun{s}.
Thus, we sample about five turn-over timescales for the hotter and almost
twenty for the cooler component.

The comparison of 3D vs. 1D models depends on which particular 1D model
is chosen. We compared each of our 3D models (hereafter denoted as
$\langle$3D$\rangle$, obtained from the mean temperature and pressure structure
of the full 3D model), to a corresponding standard
hydrostatic 1D model atmosphere (hereafter denoted as 1D$_{\rm LHD}$). The
$\langle$3D$\rangle$ model is a temporal and horizontal average of the 3D
structure over surfaces of equal (Rosseland) optical depth. It is only dependent
on the particular way the 3D model is averaged.  

The 1D$_{\rm LHD}$ model is calculated with a 1D atmosphere code called 
LHD. It assumes plane-parallel geometry and employs the same micro-physics
(equation-of-state, opacities) as \cobold. Convection is described by
mixing-length theory.  Somewhat arbitrary choices to be made
relate to the value of the mixing-length parameter, which formulation
of mixing-length theory to use, and how turbulent pressure is
treated in the momentum equation; see \citet{caffauS} for further
details. As usual, in the spectral synthesis of the 1D models a value
of the micro-turbulence has to be adopted. For 1D as well as 3D models
the spectral synthesis calculations were performed with the spectrum
synthesis code {\linfor} 
\footnote{more information on {\linfor} can be found in 
the following link:
http://www.aip.de/$\sim$mst/Linfor3D/linfor\_3D\_manual.pdf}.     

\subsection{3D corrections}

There are two main effects that distinguish 3D from 1D models,
the average temperature profile and the horizontal temperature fluctuations.
We quantify the contribution of both effects by introducing
the 3D correction as: {\ftac}.

The average temperature profile provided by a hydrodynamical simulation is
different from that of a 1D atmosphere assuming radiative equilibrium. This
effect is shown in Fig.\ref{mod3d}, where the 3D average temperature profile,
plotted as a function of the pressure, is compared to the profiles from
1D$_{\rm LHD}$ and MARCS models. As evident from the plot, the
$\langle$3D$\rangle$ temperature profile is cooler than both 1D models in the
outer photospheric layers for both of the stars. This often-encountered
effect in metal-poor atmospheres \citep{asp99} is particularly important
for the oxygen abundances derived from OH molecules, as the difference is
largest precisely in the layers where these lines are formed. The result is
that the oxygen abundances become lower in the 3D formulation than in the 1D;
we quantify this effect through the 3D correction as defined above.
 
\begin{figure}
\centering
\includegraphics[clip=true,width=9cm]{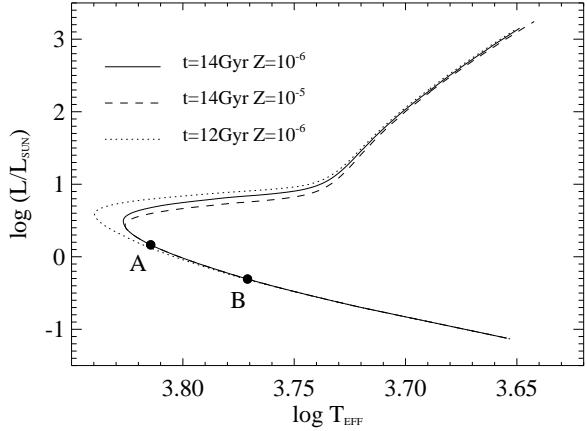}
\caption{The two components of CS 22876--032 on a 14-Gyr isochrone by 
Chieffi \& Limongi (priv. comm.) for $Z=10^{-6}$ (solid line). 
For comparison, isochrones for 14 Gyr and $Z=10^{-5}$ (dashed) and 12 Gyr and
$Z=10^{-6}$ (dotted) are also shown.} 
\label{iso}
\end{figure}

\section{Chemical analysis}

\subsection{Stellar parameters\label{satm}}

The atmospheric parameters of each star in the CS 22876--032 system were estimated from the
photometric data available in \citet{nor93} and \citet{pre91}, from whom we
adopt $V=12.84$, $B-V=0.397$, $U-B=-0.255$ with uncertainties of 0.01, 0.02 and
0.01 respectively. We adopt $E(B-V)=0.00 \pm 0.01$ from \citet{nor00} and
\citet{sch93,sch96}. We also extracted, from the 2MASS\footnote{The Two
Micron All Sky Survey is a joint project of the University of
Massachusetts and the Infrared Processing and Analysis
Center/California Institute of Technology, funded by the National
Aeronautics and Space Administration and the National Science
Foundation.} database, $K=11.503\pm0.035$ and $J=11.800 \pm 0.041$. The 
equations derived by \citet{carpenter} to transform from
2MASS magnitudes to the homogenised photometric system of
\citet{bab88} were then applied. 

From the above information it is possible to estimate the 
reddening-corrected colours $(U-B)_0$, $(B-V)_0$, $(V-K)_0$ and $(J-K)_0$, 
which we use to derive the parameters of both components of the
binary by comparing with theoretical isochrones. We have chosen the
isochrone of Chieffi \& Limongi (private communication) for 14 Gyr and
metallicity $Z=10^{-6}$, from which one can compute composite colours
from pairs of two models that lie on that isochrone. Thus, the stellar parameters 
can be derived from the best fit to the observed colours that also 
satisfy the mass ratio determined from the orbital solution. 

We note that the isochrones of Chieffi \& Limongi use the colour
transformations based on ATLAS model atmospheres, for this
temperature range, and the synthetic
colours of \citet{bcp98}. 

The result is shown in Fig.~\ref{iso}, which corresponds
to a primary star with \teffo$_{,A}=6500$ K and \loggo$_{A}=4.4$ and a
secondary star with \teffo$_{,B}=5900$ K and \loggo$_{B}=4.6$. We
checked that a change of $\Delta$Age$\;=-2$ Gyr translates into a
change of +25 K and $-$0.01 dex for the \teffo$_{,A,B}$ and
\loggo$_{A,B}$ respectively, whereas a variation of the metallicity of
$\Delta\log Z=+1$ dex does not have any impact on the resulting
stellar parameters (see Fig.~\ref{iso}). 

\begin{figure}
\centering
\includegraphics[clip=true,width=8.5cm]{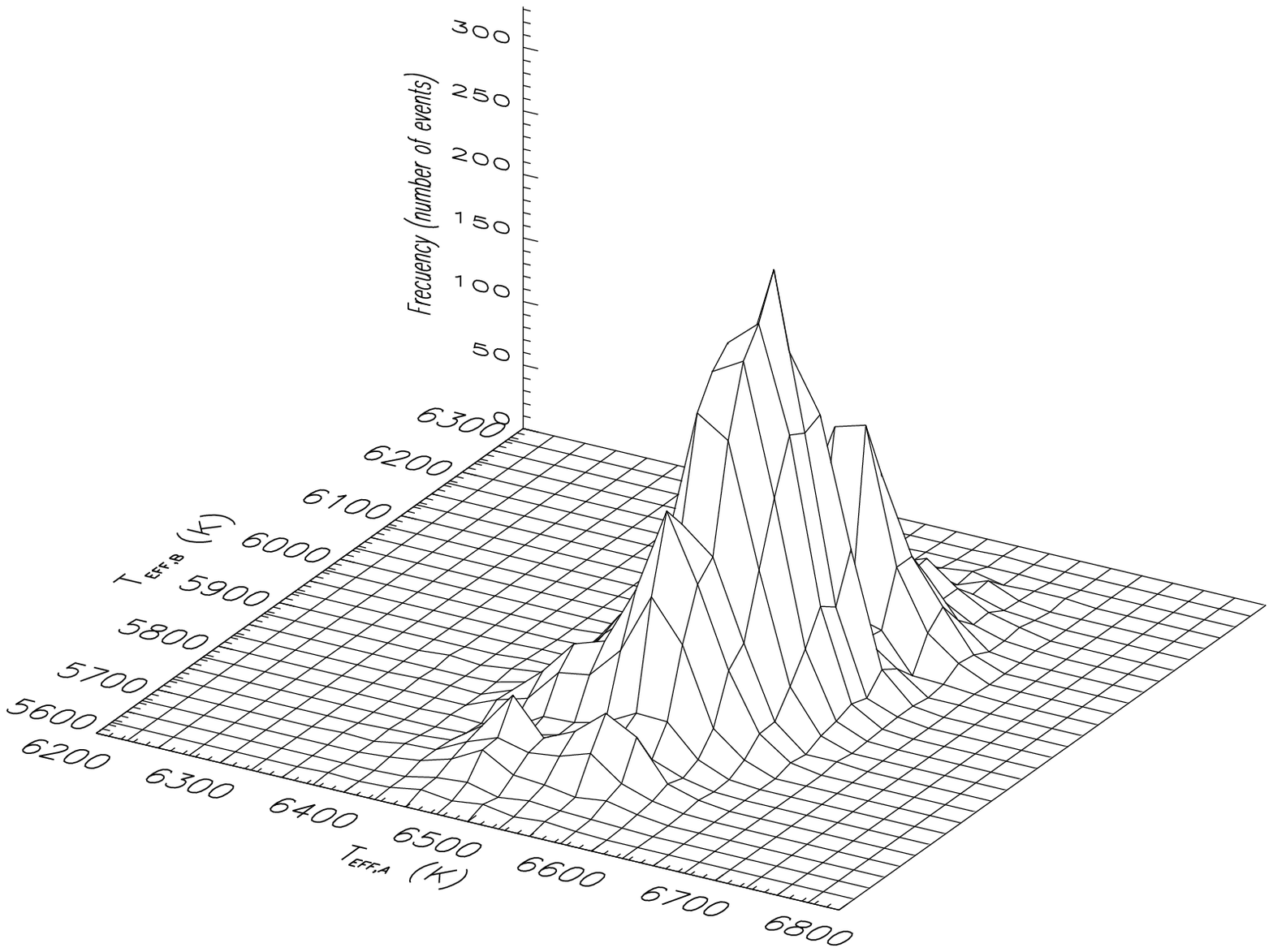}
\includegraphics[clip=true,width=8.5cm]{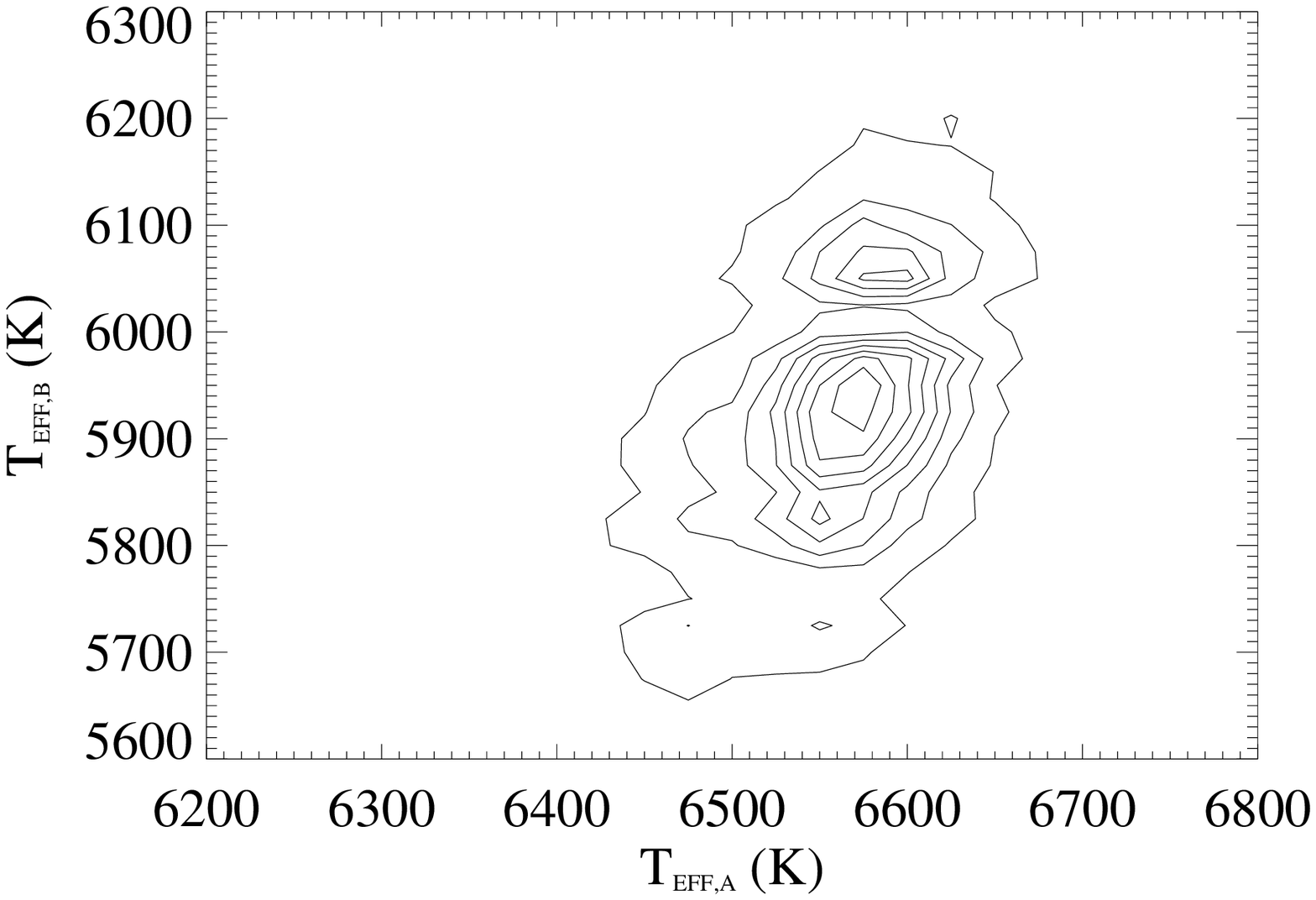}
\caption{Distribution of effective temperatures for CS 22876--032 A and B
obtained by Monte Carlo simulations, comparing the observed colours with 
those from the isochrone in Fig.~\ref{iso} for the observed mass ratio. The lowest contour encloses 95.4\% of the 10,000 simulations.}   
\label{tAB}
\end{figure}

\begin{figure}
\centering
\includegraphics[clip=true,width=8.5cm]{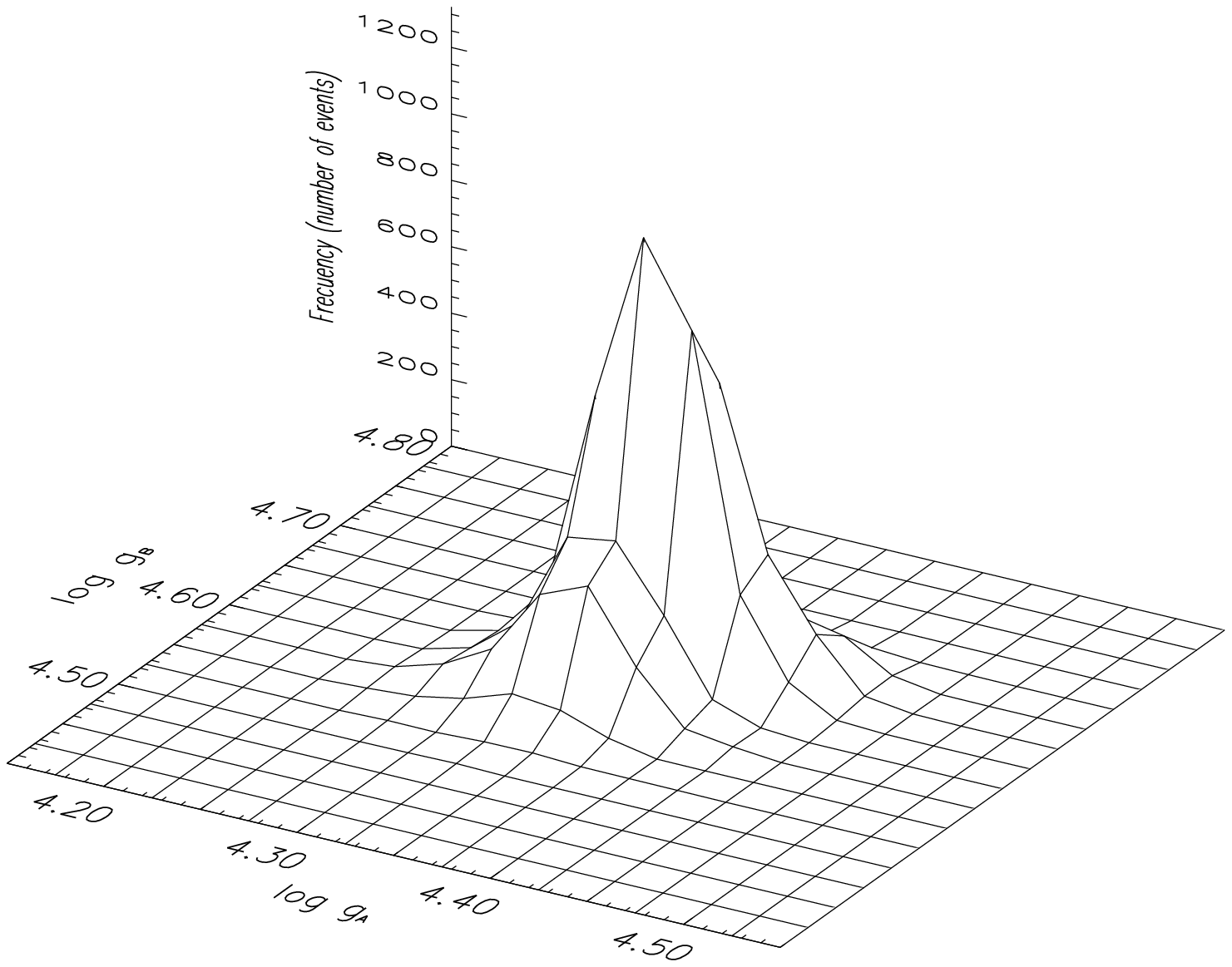}
\includegraphics[clip=true,width=8.5cm]{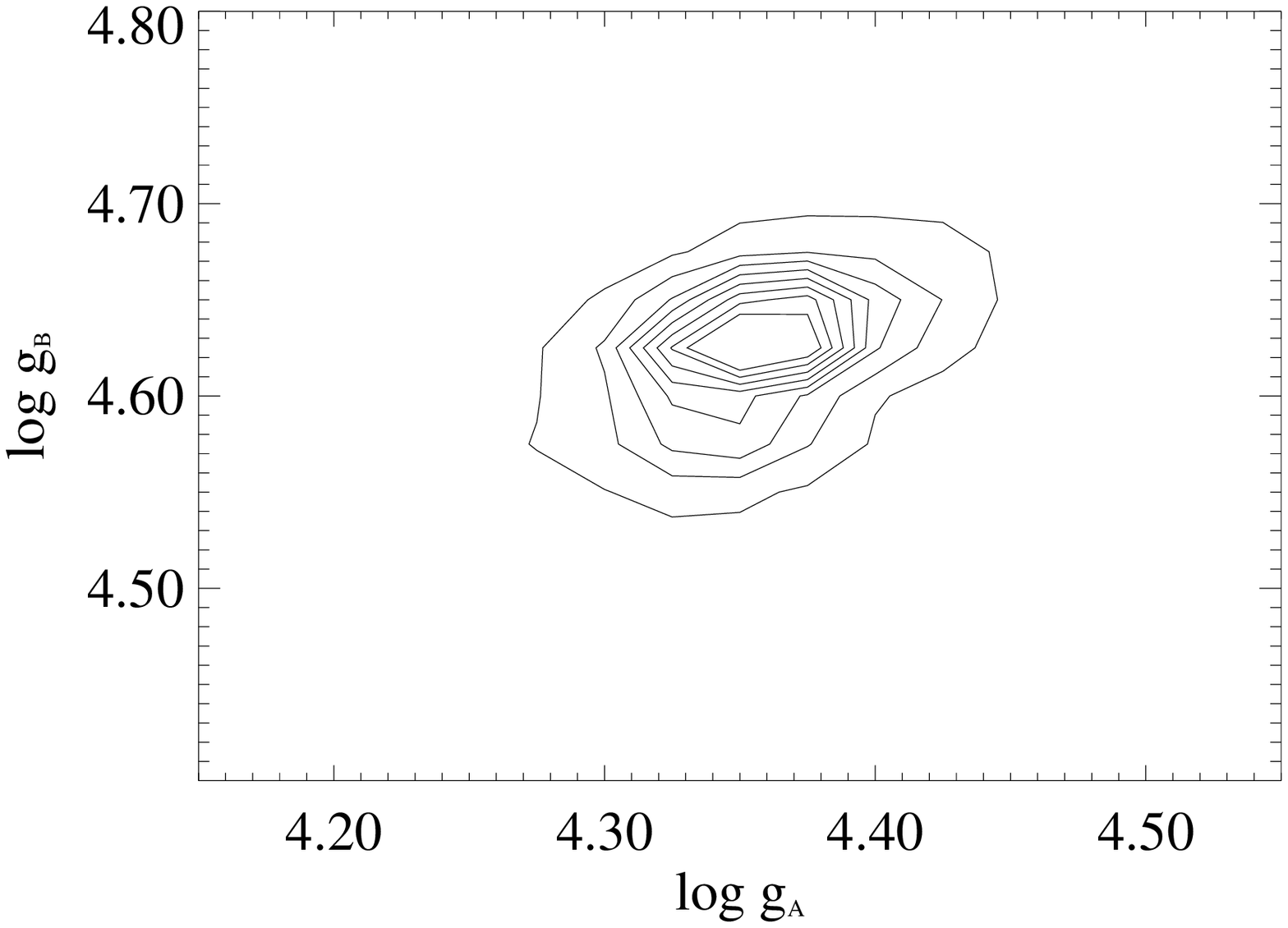}
\caption{Same as Fig.~\ref{tAB}, for the surface gravities.}   
\label{gAB}
\end{figure}

The uncertainties of the stellar parameters were estimated using Monte Carlo
techniques. We injected noise in the seven observed quantities, $V$, $B-V$,
$U-B$, $K$, $J$, $E(B-V)$, and $M_B/M_A$ following Gaussian distributions with
standard deviations equal to the errors of these quantities. From these
distributions we computed a set of five variables, $(U-B)_0$, $(B-V)_0$, $(V-K)
_0$, $(J-K)_0$, and $M_B/M_A$ for the 10,000 samples. Then we found the best fit
to each of these set of variables via a $\chi^2$ minimisation, defining
$\chi^2=\sum_{i=1}^{5}(f_{i,\rm obs}-f_{i,\rm mod})^2$, $f_{i,\rm obs}$ being
the ``observed'' value for each Monte Carlo simulation and $f_{i,\rm mod}$ the
value extracted from two pairs of points in the theoretical isochrone. 

\begin{figure}
\centering
\includegraphics[clip=true,height=9cm,angle=90]{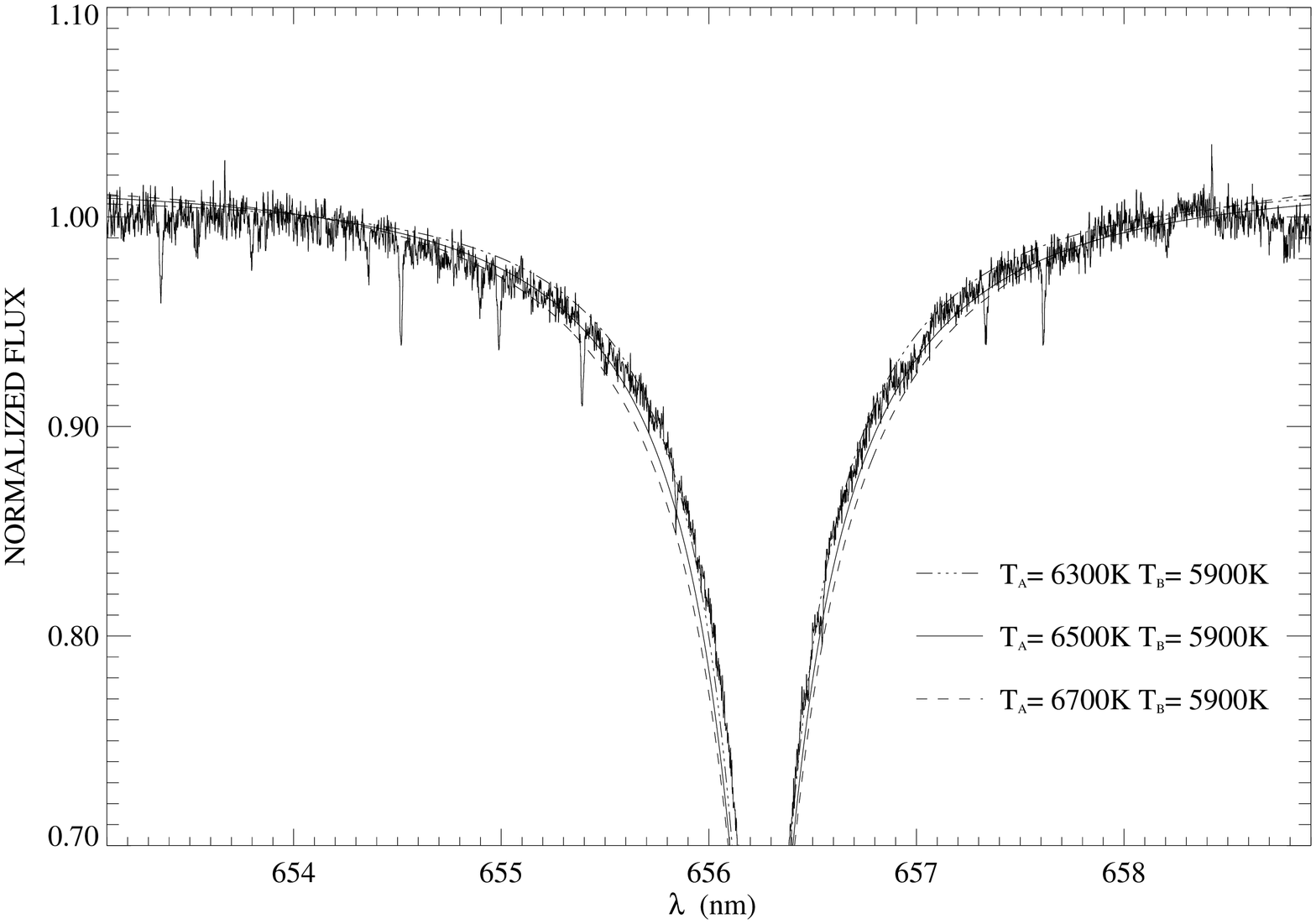}
\caption{Computed $H_\alpha$ profiles for three sets of effective 
temperatures, 6300+5900, 6500+5900 and 6700+5900, and normalised to 
the level of the observed UVES spectrum at 645.0 nm.} 
\label{halpha}
\end{figure}

\begin{table*}[ht]
\centering
\caption[]{Element abundances of CS 22876-032. The Solar O abundance is 
adopted from the 3D measurement of atomic lines of \citet{LS07};  the other Solar 
abundances from \citet{gas00}. [X/H] and [X/Fe] are LTE values; [X/Fe] 
refers to \ion{Fe}{i} for OH and neutral species, to \ion{Fe}{ii} for 
ionised species. The 3D abundance corrections $\Delta_{\rm 3D-1D}$ were 
determined with {\scshape turbospectrum} from the $\langle$3D$\rangle$ and 1D models, except for O and Li, where a full 3D analysis was performed  
and we give two values: The abundance correction {\tac} (``$\langle$3D$\rangle$'') and ``1D'' = {\ftac} (see text). 
$\sigma$ is the standard deviation of the results from the $n$ lines 
(next column; if $n=1$, the wavelength of the line in nm is given). 
See text for details on the measurement of Li, O, Sc, and Co.}             
\centering          
\scriptsize{
\begin{tabular}{lrrrrrrrrrrr}     
\noalign{\smallskip}
\noalign{\smallskip}
\noalign{\smallskip}
\hline
\hline       
\noalign{\smallskip}
Species & $\log \epsilon(\mathrm{X})_{\odot}$ & $[{\rm X}/{\rm H}]_{\rm A}$ 
& $[{\rm X}/{\rm Fe}]_{\rm A}$ & $\Delta_{\rm 3D-1D, A}$ & 
$\sigma_{\rm A}$ & $n_{\rm A}$ & $[{\rm X}/{\rm H}]_{\rm B}$ & 
$[{\rm X}/{\rm Fe}]_{\rm B}$
& $\Delta_{\rm 3D-1D, B}$ & $\sigma_{\rm B}$ & $n_{\rm B}$ \\    
\noalign{\smallskip}
\hline
\noalign{\smallskip}
\noalign{\smallskip}
\ion{Li}{i}& -- & 2.22 & -- & $^{\rm \langle3D\rangle: -0.17}_{\rm 1D: -0.19}$  & 0.01 & 670.8 & 1.75 & -- & $^{\rm \langle3D\rangle: -0.02}_{\rm 1D: -0.29}$ & 0.04 & 670.8 \\ 
\noalign{\smallskip}
O (OH) & 8.72 & -1.52 & 2.14 & $^{\rm \langle3D\rangle: -0.64}_{\rm 1D: -1.49}$ & 0.04 & 4  & -1.75 & 1.81 & $^{\rm \langle3D\rangle: 0.00}_{\rm 1D: -0.92}$ & 0.09 & 9 \\ 
\noalign{\smallskip}
\ion{Na}{i}  & 6.33 & -3.69 & -0.03 & -0.01 & 0.03 & 2    & -3.79 & -0.22 &  0.02 & 0.05 & 589.5 \\ 
\ion{Mg}{i}  & 7.58 & -3.11 &  0.55 &  0    & 0.18 & 9    & -3.14 &  0.43 &  0.09 & 0.24 & 9 \\ 
\ion{Al}{i}  & 6.47 & -3.89 & -0.23 & -0.02 & 0.22 & 2    & -3.75 & -0.18 &  0.04 & 0.22 & 2 \\ 
\ion{Si}{i}  & 7.55 & -3.75 & -0.09 & -0.01 & 0.02 & 390.5 & -3.48 &  0.09 &  0.07 & 0.04 & 390.5 \\ 
\ion{Ca}{i}  & 6.36 & -3.65 &  0.01 & -0.07 & 0.02 & 422.6 & -3.68 & -0.11 & -0.04 & 0.04 & 422.6 \\ 
\ion{Ca}{ii} & 6.36 & -3.68 & -0.18 &  0.10 & 0.14 & 3    & -3.81 & -0.54 &  0.23 & 0.11 & 2 \\ 
\ion{Sc}{ii} & 3.17 & -3.73 & -0.22 &  0    & 0.18 & 2    & -3.57 & -0.30 &  0.04 & 0.11 & 2 \\ 
\ion{Ti}{ii} & 5.02 & -3.25 &  0.25 & -0.01 & 0.34 & 19   & -3.45 & -0.18 &  0.09 & 0.23 & 12 \\ 
\ion{Cr}{i}  & 5.67 & -3.79 & -0.13 & -0.07 & 0.12 & 5    & -3.92 & -0.35 & -0.08 & 0.35 & 4 \\ 
\ion{Cr}{ii} & 5.67 & -3.38 &  0.12 &  0.08 & 0.10 & 313.2 & -3.30 & -0.03 &  0.24 & 0.30 & 313.2 \\ 
\ion{Mn}{ii} & 5.39 & -4.01 & -0.51 &  0.09 & 0.17 & 344.1 & -3.86 & -0.59 &  0.14 & 0.37 & 344.1 \\ 
\ion{Fe}{i}  & 7.50 & -3.66 &  0    & -0.12 & 0.12 & 38   & -3.57 &  0    & -0.07 & 0.21 & 37 \\ 
\ion{Fe}{ii} & 7.50 & -3.51 &  0    &  0.09 & 0.25 & 13   & -3.27 &  0    &  0.18 & 0.31 & 9 \\ 
\ion{Co}{i}  & 4.92 & -2.91 &  0.75 & -0.13 & 0.07 & 7   & -2.92 &  0.65 & -0.07 & 0.08 & 4 \\ 
\ion{Ni}{i}  & 6.25 & -3.49 &  0.17 & -0.19 & 0.18 & 12   & -3.45 &  0.12 & -0.08 & 0.16 & 12 \\ 
\noalign{\smallskip}
\hline
\hline
\noalign{\smallskip}
\noalign{\smallskip}
\noalign{\smallskip}
\end{tabular}
}
\rm
\label{tbl7}      
\end{table*}

The results of these simulations for the effective temperature and surface
gravity of both components are shown in Figs.~\ref{tAB} and~\ref{gAB}
respectively. The lowest contour encloses roughly 95.4\% of the 10,000
Monte Carlo events, analogous to $2\sigma$ for a normal distribution. From these
simulations we adopted an error, at the $2\sigma$ level, of $\Delta$\teffo$_{,
A}=100$ K and $\Delta$\teffo$_{,B}=150$ K for the effective temperature, and
$\Delta$\loggo$_{A,B}=0.1$ dex for the surface gravity. 

For single stars,  the wings of H$\alpha$ is also a very good
temperature indicator \citep{cay88,fuh93,vam96,bar02}. Adopting the
broadening theory of \citet{bar00}, we computed H$\alpha$ profiles for
several effective temperatures, using {\scshape turbospectrum}.
Unfortunately, all our UVES spectra were obtained near maximum line
separation, and the velocity difference is of the order of $\sim 30$
\kms, which precludes separation of the individual H$\alpha$ profiles.
Therefore, we had to compute a composite spectrum in order to match
the observed profile. 

Fig.~\ref{halpha} compares this synthetic H$\alpha$ profile with the 
observed profile for several combinations of effective temperatures
\teffo$_{,A}$ + \teffo$_{,B}$. We did not vary \teffo$_{,B}$ because
the H$\alpha$ absorption line of the cooler star B is weaker and
severely veiled by the flux of star A, so the combined profile is not
sensitive to changes in \teffo$_{,B}$. This comparison seems to
confirm our estimate of the effective temperature from the colours,
suggesting as well that it is on the same scale as the H$\alpha$-based
temperatures.   

\begin{table*}
\centering
\caption[]{Abundance errors in CS 22876--032. $\Delta_{T_\mathrm{eff}}$, 
$\Delta_{\log g}$, and $\Delta_{\xi}$ are the abundance changes caused 
by changes in $T_\mathrm{eff}$ of 100K (A) or 150K (B), in $\log g$ by 
0.1 dex, and by 0.5 \kms in the microturbulence velocity $\xi$. 
Other column headings and comments as in Table~\ref{tbl7}.}             
\centering          
\scriptsize{
\begin{tabular}{lrrrrrrrrrrr}     
\noalign{\smallskip}
\noalign{\smallskip}
\noalign{\smallskip}
\hline
\hline       
\noalign{\smallskip}
Specie & $\log \epsilon(\mathrm{X})_{\odot}$ & $[{\rm X}/{\rm H}]_{\rm A}$ &
$\Delta_{{\rm A,}T_\mathrm{eff}}$ & $\Delta_{{\rm A,}\log g}$ &
$\Delta_{{\rm A,}\xi}$ & $n_{\rm A}$ & $[{\rm X}/{\rm H}]_{\rm B}$ &
$\Delta_{{\rm B,}T_\mathrm{eff}}$ & $\Delta_{{\rm  B,}\log g}$ & $\Delta_{{\rm B,}\xi}$ & $n_{\rm B}$ \\     
\noalign{\smallskip}
\hline
\noalign{\smallskip}
\noalign{\smallskip}
\ion{Li}{i} & -- & 2.22 & 0.06 & 0 & 0 & 670.8 & 1.75 & 0.08 & 0 & 0 & 670.8 \\ 
O (OH) & 8.72 & -1.52 & 0.13 & -0.04 & 0.01 & 4  & -1.75 & 0.23 & 0.02 & 0.04 & 9 \\ 
\ion{Na}{i}  & 6.33 & -3.69 & 0.05 &  0    & -0.01 & 2    & -3.79 &  0.07 &  0    & -0.01 & 589.5 \\ 
\ion{Mg}{i}  & 7.58 & -3.11 & 0.04 &  0    & -0.05 & 9    & -3.14 &  0.04 &  0    & -0.04 & 9 \\ 
\ion{Al}{i}  & 6.47 & -3.89 & 0.06 &  0    & -0.01 & 2    & -3.75 &  0.06 &  0.02 & -0.03 & 2 \\ 
\ion{Si}{i}  & 7.55 & -3.75 & 0.06 &  0.01 & -0.01 & 390.5 & -3.48 &  0.04 &  0.02 & -0.05 & 390.5 \\ 
\ion{Ca}{i}  & 6.36 & -3.65 & 0.06 &  0    & -0.06 & 422.6 & -3.68 &  0.04 &  0.01 & -0.11 & 422.6 \\ 
\ion{Ca}{ii} & 6.36 & -3.68 & 0.01 &  0.03 & -0.06 & 3    & -3.81 & -0.05 &  0.05 & -0.06 & 2 \\ 
\ion{Sc}{ii} & 3.17 & -3.73 & 0.04 &  0.03 & -0.01 & 2    & -3.57 &  0.04 &  0.05 & -0.01 & 2 \\ 
\ion{Ti}{ii} & 5.02 & -3.25 & 0.05 &  0.03 & -0.04 & 19   & -3.45 &  0.01 &  0.05 & -0.09 & 12 \\ 
\ion{Cr}{i}  & 5.67 & -3.79 & 0.08 &  0    & -0.01 & 5    & -3.92 &  0.10 & 0.02 & -0.02 & 4 \\ 
\ion{Cr}{ii} & 5.67 & -3.38 & 0.03 &  0.03 & -0.07 & 313.2 & -3.30 & -0.05 &  0.05 & -0.08 & 313.2 \\ 
\ion{Mn}{ii} & 5.39 & -4.01 & 0.04 &  0.03 & -0.01 & 344.1 & -3.86 & -0.14 &  0.04 & -0.01 & 344.1 \\ 
\ion{Fe}{i}  & 7.50 & -3.66 & 0.09 & -0.01 & -0.05 & 38   & -3.57 &  0.06 & -0.02 & -0.12 & 37 \\ 
\ion{Fe}{ii} & 7.50 & -3.51 & 0.03 &  0.02 & -0.02 & 13   & -3.27 & -0.02 &  0.03 & -0.06 & 9 \\ 
\ion{Co}{i}  & 4.92 & -2.91 & 0.09 &  0    & -0.01 & 7   & -2.92 &  0.09 &  0.01 & -0.07 & 4 \\ 
\ion{Ni}{i}  & 6.25 & -3.49 & 0.09 &  0    & -0.05 & 12   & -3.45 &  0.07 &  0.01 & -0.13 & 12 \\ 
\noalign{\smallskip}
\hline
\hline
\noalign{\smallskip}
\noalign{\smallskip}
\noalign{\smallskip}
\end{tabular}
}
\normalsize
\rm
\label{tbl8}  
\end{table*}

\subsection{Stellar elemental abundances\label{sabu}}
 
Most of the elemental abundances were determined from equivalent width
measurements of selected unblended lines. These were made with an automatic
line-fitting procedure based on the algorithms of
\citet{cha95}, which performs both line detection and gaussian fits to
unblended lines. The implementation is the same as described in 
\citet{fra03}. The equivalent widths were then corrected for the
appropriate veiling factors and provided as input to {\scshape
turbospectrum} to determine the abundances. The detailed line-by-line
abundances, together with observed EWs, veiling factors, and atomic
data for both components can be found in Table~\ref{BIGtable}. 

The mean abundances for each element, listed in
Table~\ref{tbl7}, are computed using the adopted stellar
parameters derived in \S~\ref{satm} and a microturbulence of $\xi_{\rm
t}=1.5$ \kmso.
We have only performed full 3D computations with \linfor\ for the Li
doublet and the OH lines. A full 3D analysis for the hundreds of lines
involved in this work is a considerable computational task, well
beyond the scope of the present paper. However, we used the
$\langle$3D$\rangle$ models as input to {\scshape turbospectrum} to
estimate the expected corrections due to the different average
temperature profiles of the 3D models.
From the full 3D spectrum synthesis performed for Li and OH, we expect
this to be close to the {\it true} 3D correction for star B, while we 
expect significant contributions from the
temperature fluctuations in star A. In the following we
refer to \tact\ to represent the difference between the abundance
found by {\scshape turbospectrum} using a $\langle$3D$\rangle$ model
and that found using a MARCS 1D model. This is to distinguish them from
the {\it true} 3D corrections.

\subsubsection{Veiling corrections\label{veil}}

In a double-lined spectrum the strength of each spectral line, in 
particular those of the fainter component, is reduced by veiling from 
the continuum flux of the other star. Thus, the measured equivalent widths
for each spectral component must be corrected for this veling effect,
in order to obtain the intrinsic
values. The corrected equivalent widths can be estimated by
multiplying the observed EWs by veiling factors, $f_{\lambda,i}$, 
which solve the equation $1/f_{\lambda,A}+1/f_{\lambda,B}=1.0$ and
where $f_{\lambda,B}/f_{\lambda,A}$ is the primary-to-secondary
luminosity ratio. The values $f_{\lambda,i}$ are wavelength dependent
and can be estimated theoretically by computing the flux of each
stellar component, taking into account the ratio of the stellar radii.
Thus, the luminosity ratio can be expressed as
$f_{\lambda,B}/f_{\lambda,A}=F_{\lambda,A}/F_{\lambda,B}\times(R_A/R_B)^2$,
where $F_{\lambda,i}$ and $R_i$ are the flux and radius of each star.

For consistency with the isochrone colours, we use 
version 9 of the ATLAS code
\citep{1993KurCD..13.....K,2005MSAIS...8...14K} in its Linux version
\citep{2004MSAIS...5...93S,2005MSAIS...8...61S} to compute model 
atmospheres and fluxes for each star by adopting the \teff and \logg
derived in \S~\ref{satm}. We used the ``NEW'' Opacity Distribution
Functions \citep{2003IAUS..210P.A20C}, with 1 \kms\ micro-turbulence, a
mixing-length parameter~\mlp\ of 1.25, and no overshooting. The
formulation of the mixing length is different between MARCS and ATLAS;
that used in our MARCS models roughly corresponds to
~\mlp\ $\sim 1.1$ in the ATLAS formulation.

The veiling factors thus
computed differ from the MARCS+{\scshape turbospectrum} ones on
average by 0.02 which, for a line on the linear part of the curve of
growth, translates into a difference in abundance of 0.009 dex, which 
is negligible within the accuracy of our analysis. This allowed us to
compute quickly veiling factors for different combinations of the
parameters of the two stars, which were then used to estimate the
associated uncertainties. For consistency, we used the ATLAS veiling
factors throughout. 

The ratio of the stellar radii was extracted from the theoretical
isochrone, being $R_A/R_B=1.4$. The derived veiling factors lie in the range
$f_{\lambda,A}=1.29-1.37$ ($f_{\lambda,B}=4.47-3.70$) in the
spectral region $\lambda\lambda$300.0--700.0 nm. These estimates also
compare well with those used by \citet{nor00}, although they adopted 
fixed values for large spectral regions. In particular, they used
$f_{\lambda,A}=1.28$ and $f_{\lambda,B}=4.60$ for all Fe lines between
370 and 440 nm, whereas we used $f_{\lambda,A}=1.28-1.31$ and
$f_{\lambda,B}=4.19-4.58$ in that spectral region (see
Table~\ref{BIGtable}). 

In addition to the 1D veiling factors computed for the effective
temperatures of both stars, we also calculated the
$\langle$3D$\rangle$ veiling factors using the $\langle$3D$\rangle$
atmospheric models (whose temperature structure is different from that 
of the OSMARCS 1D models). These veiling factors were adopted to
properly correct the EWs given as input to the {\scshape
turbospectrum} code to estimate the {\tact} abundance corrections (see
\S~\ref{sabu}). In addition, we note that the {\ftac}
correction (see \S~\ref{s3dm}), computed only for Li and OH lines, does
not consider different veiling factors for 1D and 3D models. In this
case, we computed only the 3D veiling factors using the continuum flux
provided by the full 3D model of each star in the spectral region
close to these lines. Thus, these 3D veiling factors were applied to
the observed EWs, and the resulting EWs were used to compute the
{\ftac} corrections reported in Table~\ref{tbl7}. 

\begin{table*}[!ht]
\centering
\caption[]{Li abundances in CS 22876-032. $A{\rm (Li)_{1D,NLTE,DC}}$ includes 
corrections for depletion by diffusion (see text).}
\begin{tabular}{lrrrrrrr}     
\noalign{\smallskip}
\noalign{\smallskip}
\noalign{\smallskip}
\hline
\hline       
\noalign{\smallskip}
Component & EW$_{\rm obs}$ (pm) & $f_{6708,\rm 1D}$ & $A {\rm (Li)_{1D}}$ &$A{\rm (Li)_{1D,NLTE}}$ & $A{\rm (Li)_{1D,NLTE,DC}}$  & 
$\Delta_{\rm 3D-\langle3D\rangle}$ & $\Delta_{\rm 3D-1D}$  \\
\noalign{\smallskip}
\hline
\hline
\noalign{\smallskip}
Star A & 1.32 & 1.36 & 2.22  & 2.18 & 2.18  & -0.17 & -0.19\\ 
Star B & 0.42 & 3.74 & 1.75  & 1.77 & 1.84  & -0.02 & -0.29\\ 
\noalign{\smallskip}
\hline
\hline
\noalign{\smallskip}
\noalign{\smallskip}
\noalign{\smallskip}
\end{tabular}
\normalsize
\rm
\label{tbl6}  
\end{table*}    

\subsubsection{Uncertainties in the abundance analysis\label{errors}}

The abundance measurements are dependent on the model parameters, i.e.
effective temperature, surface gravity and microturbulence.
However, in the analysis of a spectroscopic binary, it is not possible
to avoid the influence of the veiling factors on the error estimates.
The veiling factors depend on the effective temperatures and surface
gravities of both stars. Therefore, in order to estimate the sensitivity of an
element's abundance to a given stellar parameter, one should also estimate
how veiling factors change when one of the stellar parameters of each
star is modified. Thus, we have also computed 1D veiling factors for
four pairs of models \teffo$_{,A}$/\loggo$_{A}$, \teffo$_{B}$/\loggo$_{B}$
according to the errors of the stellar parameters (see \S~\ref{satm}), by changing
one stellar parameter and fixing the three remaining parameters. The
uncertainties on the elemental abundances due to the errors of the
different model parameters are listed in Table~\ref{tbl8}. The
uncertainty of the microturbulence was assumed to be 0.5 \kmso.   

The errors computed from the dispersion of the line measurements and the
signal-to-noise ratios are listed in Table~\ref{tbl7}. We used the
Cayrel formula \citep{cay88} to estimate the errors of the
observed EWs. Due to the high S/N of the spectra we have obtained, these errors are
typically $\la 0.1$ pm in most of the spectral region covered,
except for the blue spectrum at 310.0--320.0 nm where the S/N ratio
drops significantly. Thus, the dispersion of the measurements for elements with
lines in this range is larger than for lines above 400 nm. In Table~\ref{tbl7},
we give the larger of these two estimates for each element in each star. For Li,
the error due to the uncertainty in the continuum level has been computed from a
Monte Carlo simulation, by injecting noise corresponding to the actual S/N ratio
near the Li line in the best-fit synthetic spectrum. In each case, the S/N ratio
was estimated taking into account the corresponding veiling factors. 

\subsubsection{The iron abundance \label{met}}

We made a careful selection of 38 reliable Fe I lines in star A and
37 in B, taking into account the radial velocity separation of the two 
sets of lines in the double-lined spectrum. It is reassuring that we 
find the same [Fe/H] for both stars, within the errors. Note that in 
our analysis, in contrast with \citet{nor00}, this was not imposed 
{\em a priori}, and thus supports our determination of atmospheric
parameters. Our value of [Fe/H] is very close to that by \citet{nor00} 
despite the very different effective temperatures adopted. This results from
the fact that different effective temperatures also imply different veiling factors, 
which must be factored in. 

Reviewing the \tact\
corrections, we see that [Fe/H] is reduced in 3D, and the difference 
between the two stars increases. We expect the true 3D corrections of the 
primary star to be largely due to horizontal temperature fluctuations, 
therefore probably larger than our \tact\ correction, and we predict that a 
full 3D-LTE synthesis will not improve the agreement of [Fe/H] between the 
two stars. Our best estimate of [Fe/H] of the system is still the \tact\ 
corrected value for star A: [Fe/H]=--3.78, which might be further reduced 
by a full 3D-LTE synthesis. This confirms that the stars in CS 22876--032 
are the most metal-poor dwarfs known to date. 

We note that ionisation equilibrium is not achieved in either star. In both, the
abundance derived from the \ion{Fe}{ii} lines is larger, for star B by a factor
of two. We note, however, that the \ion{Fe}{ii} abundance shows a very large
scatter in both A and B (0.25 and 0.31 dex, respectively) so that, within
errors, the \ion{Fe}{i} and \ion{Fe}{ii} abundances remain compatible. The
number of \ion{Fe}{ii} lines measured is very large for stars of this
metallicity (13 in star A, 9 in B). However. all the \ion{Fe}{ii} lines are
weak and the majority of them are in the UV range, were the S/N ratio of our
spectra drops dramatically. 

The \tact\ corrections for \ion{Fe}{ii} are in the {\em opposite} direction of to those 
for \ion{Fe}{i}, making the ionisation imbalance worse. The different
signs of the \tact\ corrections for neutral and ionised species are 
due to the different ionisation structure of the $\langle3D\rangle$ and
1D models. We note that for the metal-poor subgiant HD 140283,
\citet{Shc05} have performed NLTE computations for \ion{Fe}{i} and
\ion{Fe}{ii}, using a single snapshot of a hydrodynamical simulation, 
and found 3D-NLTE corrections of +0.6 for \ion{Fe}{i} and +0.4 for
\ion{Fe}{ii}. To the extent that these computations can be considered
representative of the stars in CS 22876--032, we expect that a full 
3D-NLTE analysis might achieve a better ionisation balance for iron. 

\begin{figure}
\centering
\includegraphics[clip=true,height=9cm,angle=90]{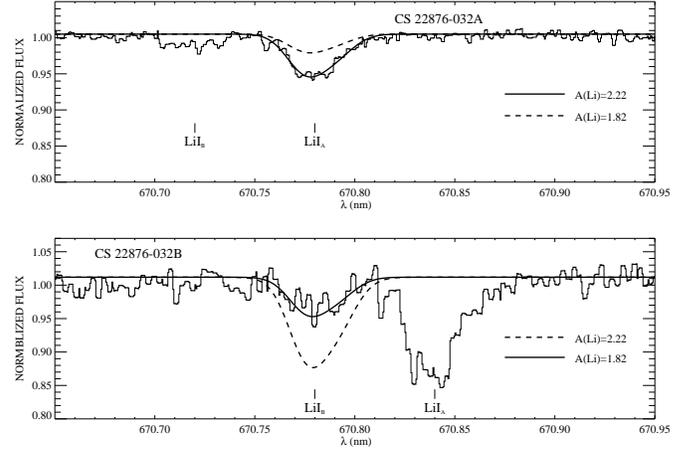}
\caption{Synthetic spectral fits to the Li line in the co-added UVES 
spectrum of star A (top) and B (bottom). The 
observed spectra have been corrected for veiling (see
Table~\ref{tbl6}), so the lines appear with their intrinsic strength
in each star.} 
\label{synLi}
\end{figure}

\subsubsection{Lithium\label{li_sec}}

The high quality of our spectra allowed us to measure the Li doublet in 
both components of CS 22876-032 for the first time. Fig.~\ref{synLi} 
shows our mean spectra of the Li region, where our well-resolved spectra 
(only) have been co-aligned on the lines of star A (top) and B (bottom), 
respectively. The superposed lines of the other star appear only slightly 
diffuse, because the velocity difference between the two stars is nearly 
the same in all our UVES spectra (see Table \ref{tbl1}). The spectra in 
Fig.~\ref{synLi} have been corrected for veiling, as discussed in \S~\ref{veil}.

The Li abundance of each star was computed from two sets of spectra taken 
on different nights. We found differences of 0.01 and 0.05 dex between 
star A and B and adopted the average value of the two measurements. 
Table~\ref{tbl6} lists the observed equivalent widths and the average 
Li abundances, with and without the NLTE corrections obtained from the 
tables of \citet{carlsson}. We further correct for the effect of 
depletion as predicted by the standard isochrones of \citet{deli}. The 
correction is negligible for the hot primary, but larger for the 
secondary (see Table~\ref{tbl6}). 

It is certainly surprising that the two components appear to have a different
lithium content. The Li abundance of the primary component seems to be
consistent with that observed in other metal-poor stars, i.e., the {\em Spite
plateau} \citep{sas82a,sas82b,BM97,ryan,melendez,CP05,asp06,bon07}, modulo the
uncertainties on the temperature scales adopted over the years by the various
authors, but the secondary star definitely seems to exhibit a lower Li content.

According to the Li depletion
isochrones of \citet{deli}, if star B were 350 K cooler than we assume, 
the correction for A(Li) would be 0.6 dex. Such a change in \teff would 
also imply a slightly higher $\log g$ from the isochrone, so the model 
dependencies given in Table \ref{tbl8} would imply that A(Li) should be 
reduced by 0.17, giving a ``corrected'' Li abundance of A(Li)=2.20,
in essential agreement with A(Li) of the primary (we do not consider
variations in the temperature of star A). By changing simultaneously 
the effective temperatures of
the pair (within the range allowed by photometry), we would derive
different Li abundances for each of the components. However, we would
also change the veiling factors, and the final abundance
difference between the two stars would be roughly the same. 

We have checked whether inaccurate veiling corrections, 
especially for the fainter lines of star B, could be the cause of the 
different Li abundances. It turns out that this is impossible: bringing 
$A{\rm (Li)_B}$ into consistency with the {\em Spite plateau} would
require a doubling of the veiling correction at 670 nm; given that we
find consistent abundances for Na at 590 nm and Mg I at 880 nm, the
veiling correction cannot be off by a factor two at the intermediate
wavelength. 

For the Li lines we also performed a full 3D-LTE synthesis using \linfor\ .
The resulting corrections are listed in Table~\ref{tbl6}. Since the 3D
computation was performed in LTE, the Li abundances must not be taken
as definitive, as shown by \citet{CS00} and \citet{asp03}. It is,
however, interesting to notice that while the 3D effect in star A is
almost entirely due to the horizontal temperature fluctuations, for
star B it is almost entirely due to the cooler average temperature
profile of the 3D model. A full 3D-NLTE synthesis of Li in CS
22876--032 is beyond the scope of this paper. Full 3D-NLTE synthesis
of the Li profile in HD74000 have been addressed recently by \citet{cay07}.  

\subsubsection{Oxygen}

\begin{figure}
\centering
\includegraphics[clip=true,height=9cm,angle=90]{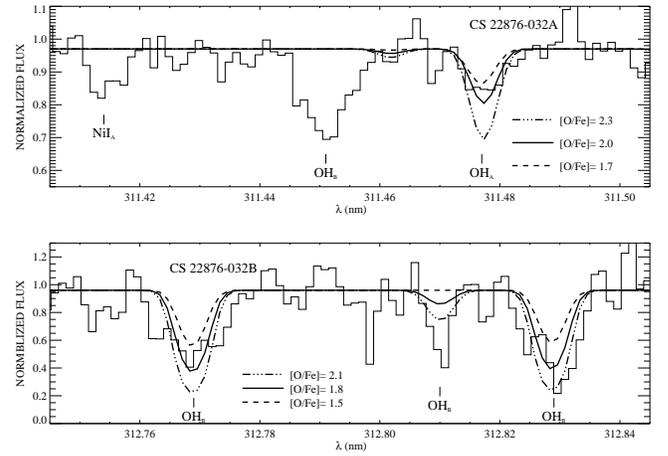}
\caption{
\label{synOH1}
Synthetic spectral fits to OH lines in the co-added UVES spectrum of 
star A (top) and B (bottom) in \mbox{CS 22876--032}. The observed spectra 
have been corrected for veiling factors of 1.30 (A) and 4.36 (B).} 
\end{figure}

\begin{figure}
\centering
\includegraphics[clip=true,height=9cm,angle=90]{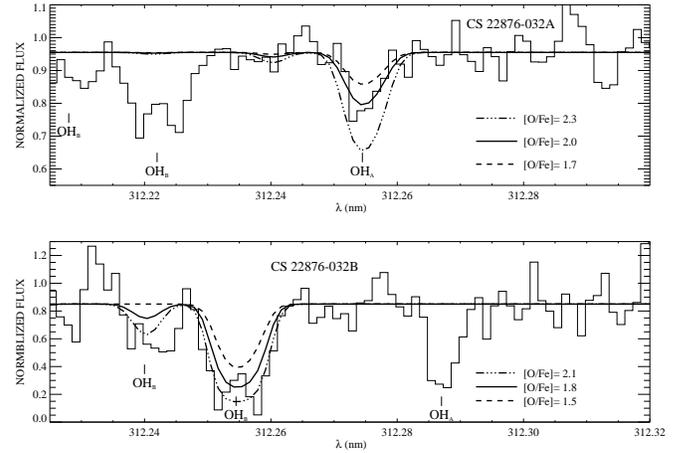}
\caption{\label{synOH2} Same as Fig.~\ref{synOH1} for another OH line.} 
\end{figure}

\begin{figure}
\centering
\includegraphics[clip=true,height=9cm,angle=90]{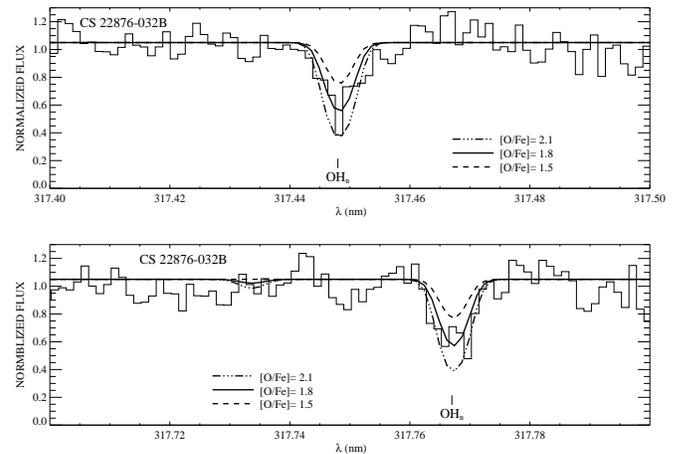}
\caption{\label{synOH3}
Synthetic spectral fits to OH lines in the co-added UVES spectrum of
star B. The observed spectra have been corrected for a veiling factor 
of 4.36.} 
\end{figure}

\begin{table}[!ht]
\centering
\caption[]{3D abundance corrections for the OH lines. [O/H] and [O/Fe] are relative to $\log \epsilon(\mathrm{O})_{\odot}=8.72$.}
\begin{tabular}{lrrrr}     
\noalign{\smallskip}
\noalign{\smallskip}
\noalign{\smallskip}
\hline
\hline
\noalign{\smallskip}
$\lambda$ (nm) & [O/H]$^{\rm a}_{\rm 1D}$ & [O/Fe]$_{\rm 1D}$ & $\Delta_{\rm 3D-\langle3D\rangle}$ & $\Delta_{\rm 3D-1D}$  \\
\noalign{\smallskip}
\hline
\noalign{\smallskip}
\multicolumn{5}{c}{CS 22876--032A} \\
\noalign{\smallskip}
\hline
\noalign{\smallskip}
\noalign{\smallskip}
311.21 & -1.48 & 2.18 & -0.633 & -1.546 \\
311.47 & -1.55 & 2.11 & -0.536 & -1.328 \\
312.26 & -1.56 & 2.10 & -0.590 & -1.420 \\
313.43 & -1.48 & 2.18 & -0.797 & -1.664 \\
\noalign{\smallskip}
\hline
\noalign{\smallskip}
\multicolumn{5}{c}{CS 22876--032B} \\
\noalign{\smallskip}
\hline
\noalign{\smallskip}
312.26 & -1.63 & 1.94 &  0.027 & -0.985 \\
312.77 & -1.85 & 1.72 &  0.014 & -0.868 \\
312.83 & -1.68 & 1.89 &  0.033 & -1.009 \\
313.03 & -1.94 & 1.63 & -0.004 & -1.045 \\
313.32 & -1.69 & 1.88 &  0.003 & -0.856 \\
313.66 & -1.75 & 1.82 & -0.015 & -0.849 \\
314.30 & -1.80 & 1.77 & -0.011 & -0.815 \\
317.45 & -1.75 & 1.82 & -0.017 & -0.916 \\
317.77 & -1.70 & 1.87 & -0.009 & -0.906 \\
\noalign{\smallskip}
\hline
\hline
\noalign{\smallskip}
\noalign{\smallskip}
\noalign{\smallskip}
\end{tabular}
\normalsize
\rm
\label{tbl5}  
\end{table}    

\begin{table}[!ht]
\centering
\caption[]{1D gravity effects on the Fe and O abundances}
\begin{tabular}{lrrr}     
\noalign{\smallskip}
\noalign{\smallskip}
\noalign{\smallskip}
\hline
\hline       
\noalign{\smallskip}
Species & $\Delta{\log g}$ & $\Delta{[{\rm X}/{\rm H}]_{A}}$ & $\Delta{[{\rm X}/{\rm H}]_{B}}$  \\
\noalign{\smallskip}
\hline
\noalign{\smallskip}
OH    & 0.3 & -0.12  & -0.08 \\ 
\ion{Fe}{i}   & 0.3 & -0.01  & -0.02 \\ 
\ion{Fe}{ii}  & 0.3 & 0.09  & 0.11 \\ 
\noalign{\smallskip}
\hline
\hline
\noalign{\smallskip}
\noalign{\smallskip}
\noalign{\smallskip}
\end{tabular}
\normalsize
\rm
\label{tbl4}  
\end{table}    

The oxygen abundances have been derived from UV OH lines of the (0-0)
vibrational band of the $A^2\Sigma - X^2\Pi$ electronic system. The
use of these lines for oxygen measurements in metal-poor stars was
pioneered by \citet{bessel}. We were able to measure four lines
in the primary and nine lines in the secondary. Following the extensive
surveys of OH lines in metal-poor stars by \citet{isr98,boe99,isr01},
and the controversial finding of strongly increasing [O/Fe] with
decreasing metallicity, \citet{agp01} warned of the
possible role of 3D effects on the formation of these lines. For this
reason we decided to compute {\em ad hoc} \cobold\ hydrodynamical
simulations for this system in order to correctly evaluate the 3D
effects. Our results are summarised in Table~\ref{tbl5}. Some of 
the analysed lines are shown in
Figs.~\ref{synOH1}, \ref{synOH2} and \ref{synOH3}.  

The 3D effects are clearly large and considerably different between
the two stars. \citet{agp01} attributed the 3D corrections
primarily to the different average temperature profile of the 3D
models, in particular to their extremely cool outer layers. However,
we see that for the primary star the 3D correction is almost evenly
shared between average temperature profile and horizontal temperature
fluctuations\footnote{$-$0.64 dex due to temperature fluctuations and
$-$0.85 dex due to average temperature profile}.

Our cooler 3D model atmosphere exhibits much smaller temperature
fluctuations around the mean than the hotter one.  In the cool
component the average temperature becomes so low that a substantial
amount of H$_2$ is formed in the higher photospheric layers. The
associated increase of the specific heat makes it much harder for
pressure fluctuations to introduce temperature fluctuations. Due to
the smaller temperature fluctuations the cooler 3D and
$\langle$3D$\rangle$ models provide essentially the same abundances.
The model of the primary star is hotter on average, H$_2$ molecules
are much less abundant, and temperature fluctuations are much more
pronounced. Consequently, the resulting abundances differ between 3D
and $\langle$3D$\rangle$ models. 

\begin{figure}
\centering
\includegraphics[clip=true,width=9cm]{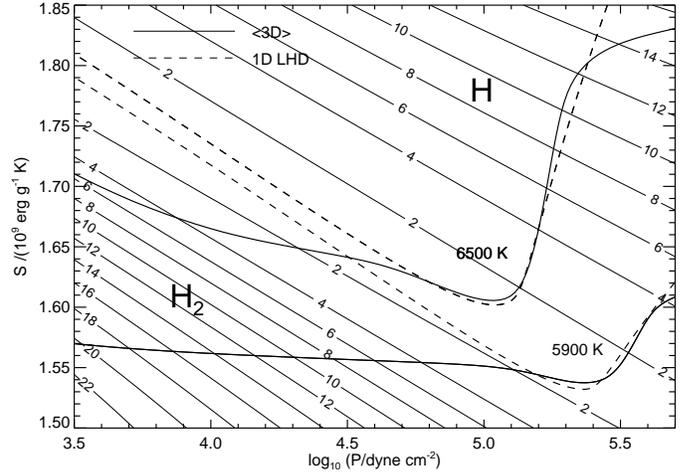}
\caption{Entropy profiles of the $\langle$3D$\rangle$ CO$^5$BOLD model 
atmospheres (solid lines), compared to 1D$_{\rm LHD}$ models (dashed) 
for stars B (5900\pun{K}) and A (6500\pun{K}). The labelled contours 
indicate the specific heat at constant pressure in units of 
$10^8$\,erg\,g$^{-1}$\,K$^{-1}$(see text).} 
\label{cp3d}
\end{figure}

Fig.~\ref{cp3d} illustrates the
situation. The 3D model exhibits a stronger cooling with
respect to the 1D$_{\rm LHD}$ model at 5900\pun{K} than is the
case at 6500\pun{K}. Over a wide pressure range the structure is 
almost adiabatic and passes through a region of substantial H$_2$
molecule formation, indicated by the rather high values of the specific
heat, suppressing temperature fluctuations in that region. The almost
adiabatic structure of the cool model also indicates that the
convective overshooting is very efficient compared to radiative
heating, and has driven the thermal structure into almost adiabatic
equilibrium. 

The large difference in the behaviour of our two models warns us that
to measure reliable abundances from OH one needs a grid of 3D models
which is fairly dense in temperature, to capture, for any metallicity,
the \teff at which H$_2$ formation sets in.

Finally, we note that the O abundance derived from OH lines is rather
sensitive to the surface gravity, as can be seen from
Table~\ref{tbl4}. An increase of gravity of 0.3 dex introduces a
decrease in O abundance of about 0.1 dex. This gravity dependence is
larger than that of \ion{Fe}{i}, however it is advisable to use
\ion{Fe}{i} to derive [O/Fe], since the gravity dependence of
\ion{Fe}{ii} is in the {\em opposite} direction. Note that the values
reported in Table~\ref{tbl4} have been estimated without revising the 
veiling factor when changing the surface gravity of the models, 
contrary to the error estimates described in \S~\ref{errors}. 

\subsubsection{Other elements}

The \ion{Be}{ii} resonance doublet at 313.0 nm is within the
wavelength range covered by our spectra. The S/N ratio in
that region is $\sim 25$, yet none of the Be lines is detected, which 
implies an upper limit for the Be abundance log(Be/H) $< -13.0$. 
As this is an order of magnitude higher than the Be abundance expected 
for these stars from the trend of Be abundance with metallicity
\citep{gilmore,ryan92,mol97,Boesgaard1999,Primas02,Boesgaard2006}, this 
result is not of deep significance. 

For the other elements, abundances were determined directly from the 
EWs, except for Sc and Co for which we used spectrum synthesis to 
take hyperfine splitting (HFS) into account. For Co we used the $A$ and $B$
factors measured by \citet{pickering}. For the \ion{Sc}{ii} 361.3\,nm 
line we used the $A$ and $B$ factors measured by \citet{HFSScII}.
In both cases we used the code {\scshape linestruc} of \citet{wahlgren} 
to compute HFS components. For the \ion{Sc}{ii} 424.6\,nm line we used
the HFS components given in Table 5 of \citet{mcw95}. The detailed
atomic data of HFS for Sc and Co are provided in Table~\ref{tblhfs}. For
Mn we used a single line, the strongest of the \ion{Mn}{ii} lines of
Mult. 3, for which hyperfine splitting is negligible according to \citet{CH04}.  

Silicon was only measured from the \ion{Si}{I} line at 390.5\,nm. In
metal-poor cool giants, the silicon abundance is derived from a line
at 410.3\,nm, since the line \ion{Si}{I} 390.5\,nm is severely blended
with CH lines \citep{cay04}. The \ion{Si}{I} at 410.3\,nm is very
weak in metal-poor dwarfs, but CH lines are so weak that the line at
390.5\,nm can be used.  

Aluminium was measured from the resonance lines
\ion{Al}{I} 394.4\,nm and \ion{Al}{I} 396.1\,nm. The \ion{Al}{I}
394.4\,nm is blended with CH lines, which are also extremely weak
in metal-poor dwarfs and were not taken into account. In fact, we
computed synthetic spectra of the CH lines at 390.5\,nm and 394.4\,nm
for the stellar parameters of both dwarfs, and the CH lines are not
visible even for [C/Fe] as large as +2 dex. However,
the 394.4\,nm line provided a significantly larger abundance than that
derived from the 396.1\,nm line in both stars. For this reason, in
Table~\ref{tbl7} we give the average Al abundance derived from both
lines with a large dispersion of $\sim 0.2$ dex. 

Using $\langle$3D$\rangle$ models, we determine the {\tact} corrections
listed in Table~\ref{tbl7}. We note that the {\tact} corrections also
take into account different veiling factors estimated using
$\langle$3D$\rangle$ models and 1D OSMARCS models. In general, the
veiling factors estimated from $\langle$3D$\rangle$ models are higher
for star B and lower for star A than those
obtained using 1D models. This effect is especially important for the
secondary star, and becomes more significant at shorter wavelengths. Taking into
account this effect, it is interesting to note that {\tact}
corrections are in general negative for neutral species and positive 
for ionised species, at least for the primary star, for which the
veiling factors are not significantly different from those estimated
using 1D models, as already noted for iron. This reflects the
different ionisation structure of 1D and $\langle$3D$\rangle$ models. 
The {\tact} corrections do not help to achieve ionisation
equilibrium for Ca, Cr and Fe, the only elements for which abundances
were determined from both neutral and ionised species. 

\section{Discussion}

\subsection{Lithium\label{li_disc}}
  
\begin{figure}
\centering
\includegraphics[clip=true,width=9cm,angle=0]{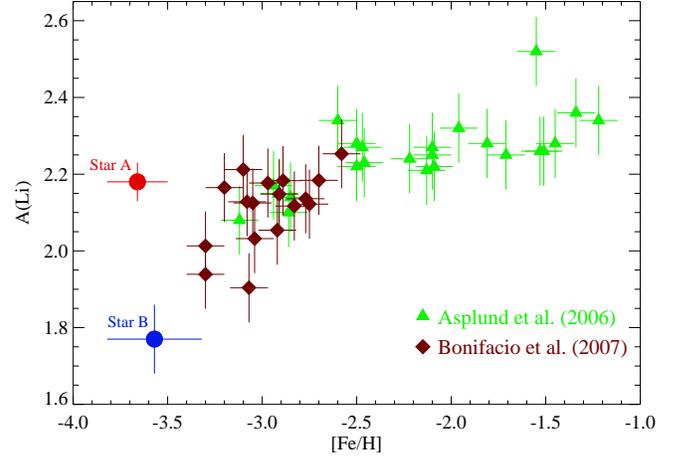}
\caption{1D-NLTE Li abundances vs. [Fe/H] for the stars in \mbox{CS 22876--032} 
(circles) and in other metal-poor dwarfs as reported by Asplund et al. 
(2006, triangles) and Bonifacio et al (2007, rhombs). The Li abundances 
by \citet{asp06} were recomputed using {\scshape turbospectrum} as 
reported by \citet{bon07}. No 3D-LTE or 3D-NLTE has been considered.}  
\label{life}
\end{figure}

\begin{figure}
\centering
\includegraphics[clip=true,width=9cm,angle=0]{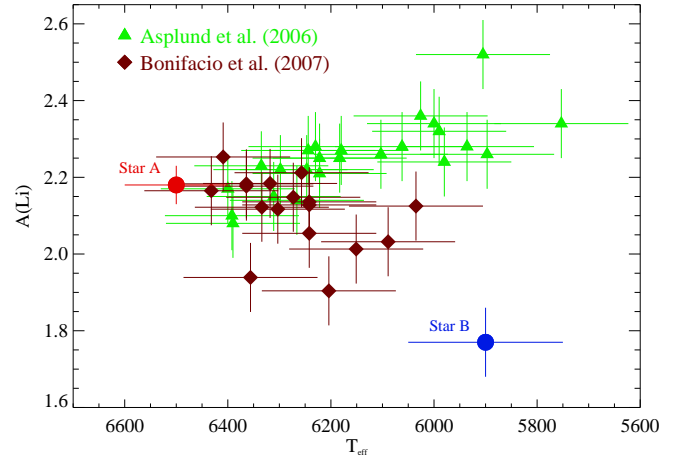}
\caption{Same as Fig.~\ref{life}, but in the Li-\teff plane.}  
\label{liteff}
\end{figure}

In Fig.~\ref{life} we show the lithium abundances for the two
components of CS 22876--032, together with our data from Paper VII,
including the data of \citet{asp06} rescaled in [Fe/H]
and A(Li) to be homogeneous with our own data. 
Star A appears to have a Li abundance at the same level as the
majority of stars with metallicity below --2.5, if anything slightly
higher. Star B appears to be far below\footnote{-0.33 dex below the
average Li abundance of the stars presented in paper VII} any of the
other measured stars. It should be noted that all the other stars in
Fig.~\ref{life} have effective temperatures determined from the wings
of H$\alpha$ using the broadening theory of \citet{bar00}, while for
CS 22876--032 they have been determined independently from
colours and isochrones. However, the reasonable agreement
between our computed H$\alpha$ profile and the observed profile shown
in Fig.~\ref{halpha} suggests that the two temperature scales are
fairly close. 

The fact that at the lowest observable metallicity, star A
remains at the level of the plateau suggests that there is no downturn
or decrease in Li abundance at the lowest metallicities. This suggests
that the slope of A(Li) with [Fe/H] which is detectable in the sample
of \citet{asp06} (but not in that of Paper VII, alone) is not real,
but rather an artifact due to the H$\alpha$ temperature scale. It is
possible that it is ultimately due to our inability to correctly model
the atmospheres of extremely metal-poor stars and the wings of Balmer
line profiles. In Paper VII we argued that the data could suggest
either a vertical drop or an increased scatter in A(Li) at the lowest
metallicities. Now, the drop seems to be ruled out by the A(Li)
measured in star A. The measurement in star B, taken at face value,
may support the idea that at metallicities below --2.5, the {\em Spite
plateau} displays a sizeable scatter. 

In section \ref{li_sec} we have pointed out how the difference of the
Li abundance between the two stars could be resolved by assuming that
the temperature of star B were 5550 K. Star B would thus be subject 
to considerable Li depletion, according to the standard
Li depletion isochrones of \citet{deli}. Such isochrones are available
only for metallicities considerably higher than that of CS 22876--032.
If, for any reason, either the lower metallicity of our system, or
inclusion of other physical phenomena, the dependence of Li depletion
on \teff is steeper than predicted by purely diffusive standard
isochrones, then to reconcile the Li abundances of the two stars, the
temperature of star B could be higher than 5550 K. Considering that
our estimated error on the effective temperature of star B is 150 K
($2\sigma$), such cooler temperatures are not totally implausible.

In our view the existence of a real scatter in Li abundances at the
lowest metallicities remains to be established beyond any
reasonable doubt. It is, nevertheless, worthwhile to discuss the
possible implications of such a scatter, if real.

We have already noted in Paper VII that by arbitrarily dividing the
sample into two sub-samples, one with metallicity below --3.0 and the
other above, the scatter of the ``higher'' metallicity sample was 0.05
dex, while the scatter of the ``lower'' metallicity sample was 0.11. 
If to the lower metallicity sample we add the two stars of \mbox{CS
22876--032}, the scatter increases only slightly, to 0.12 dex.
\citet{richard} have studied the effect of atomic diffusion  
in presence of turbulence and concluded that the observations of the
{\em Spite plateau} could be explained by starting from a primordial
A(Li), compatible with the baryonic density derived from the WMAP
experiment. The effect of atomic diffusion, countered by a suitably 
parametrised turbulence, can explain the present level and low scatter
of the {\em Spite plateau}. 

From their Li isochrones in the presence of pure atomic diffusion
\citep[Fig.~5 of ][]{richard} it is obvious that one should expect Li
abundances in the range 1.70$\le$A(Li)$\le$2.35. The sample 
constituted by the 8 stars from Paper VII with metallicity below 
--3.0 and the two stars in CS 22876--032 spans the range 1.91 - 
2.20 in A(Li). One could therefore suspect that in extremely
metal-poor stars, turbulence is lower and atomic diffusion more
efficient, causing the increased scatter in Li abundance. 
In Fig.~\ref{liteff}
we see no clear trend of A(Li) vs. \teffo. When comparing this figure
with Fig.~5 of \citet{richard} we note that the ``high'' Li
abundances of the hottest stars in the sample (among which is the primary
of CS 22876--032), preclude any clear resemblance between the two pictures.
Therefore, current pure diffusion models seem unable to explain at the same
time the behaviour of Li abundances with [Fe/H] and \teffo. 
 It is possible, however, that they may do so, after an {\em ad hoc}
parametrisation of turbulence with metallicity.

Recently \citet{k2006,k2007}
have found that the models of \citet{richard}, with 
a suitable value for the turbulent diffusion coefficient,
can explain the 0.12 dex difference in A(Li) they find 
between stars at the turn-off and on the subgiant branch stars
of the globular cluster NGC 6397.
The same authors, however point out that assigning temperatures
for the TO stars, hotter by 170 K (therefore close to the
temperatures adopted by \citealt{B02} for the TO stars
of this cluster), such a difference would vanish.

Though suggestive, the applicability of such turbulent
diffusive models remains to be proven.
The main cause of concern is the parametrisation
of the turbulent diffusive coefficient, which is
linked to a fixed temperature, and not
to the bottom of the convective zone
\citep{richard2002}. 

In order to strengthen the observational constraints on such models
and refine the estimates of the scatter and slope (or lack thereof) of
the extreme metal-poor end of the Spite Plateau, further high-quality 
spectroscopy of EMP stars and additional accurate constraints on the
effective temperatures of the whole sample are highly desirable.

\subsection{Oxygen}

\begin{figure}
\centering
\includegraphics[clip=true,width=9cm,angle=0]{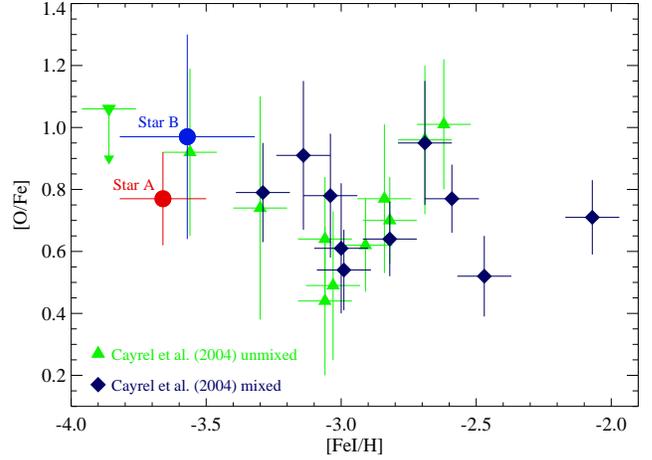}
\caption{3D [O/Fe] ratios of the stars in CS 22876--032 (circles) 
and in the metal-poor giants of Paper V. Triangles: ``unmixed'', rhombs
``mixed'' stars; downward triangle with arrow: upper limit for
CS 22172--002). 
}    
\label{ofe3dLP}
\end{figure}

Fig.~\ref{ofe3dLP} compares our 3D-LTE O abundances obtained from OH UV bands
in CS 22876--032 with the high-quality measurements of Paper V for giant
stars (mixed and unmixed), obtained from the [OI] 630nm line. We note that
we have computed the 3D-LTE [O/Fe] ratios using 
the true 3D-LTE oxygen abundances and the $\langle$3D$\rangle$
\ion{Fe}{i} abundances. We have chosen \ion{Fe}{i} rather than
\ion{Fe}{ii} abundances as the reference, because
their sensitivity to the surface gravity is similar as that of the OH
bands. The reader might wonder why we find different O abundances for
stars A and B, but they are in fact consistent within the
error bars, which mainly reflect the uncertainties in the effective
temperature (see Table~\ref{tbl8}). 

The abundances derived from [OI]
are the raw values obtained from 1D-LTE model atmospheres. One worry is
the possible effects of granulation on the abundances derived from the
[OI] line in giants. In a recent paper, \citet{collet} investigated the
3D effects in giant stars and found very small corrections
for the [OI] lines for metallicities down to --2.0, but sizeable (almost 0.2
dex) downward corrections for [O/Fe]) at metallicity --3.0 (their Fig.
13 and Table 3). If we apply the corrections interpolated and
extrapolated from Table~3 of \citet{collet} to the giants of Paper
V, the [O/Fe] decreases in all the giants and the good agreement
between CS 22876--032 and the giants no longer holds. The mean
[O/Fe] of the giants would be 0.51, while that of
CS~22876--032 is 0.87. If instead we take the measurements of Paper V
at face value, the mean [O/Fe] is 0.72. 

At present we have no full 3D
models for giant stars, however, from a few snapshots for models of
\teff = 4900, log g = 2.0 and metallicity --2.0 and --3.0 we find
little difference at the two metallicities, and very small 3D
corrections. The \ftac\  abundance correction amounts to --0.006 dex at
metallicity --2.0 and --0.037 dex at --3.0, which is negligible. For our
\cobold\ models of giants the difference between the mean 3D temperature
structure and a corresponding 1D$_{\rm LHD}$ (\mlp = 1.0) model is not large. In
particular we do not find the substantial cooling of the highest photospheric
layers at metallicity --3.0, which
\citet{collet} find (see their Fig.~1). Moreover, the mean 3D temperature
profile is slightly hotter than the 1D$_{\rm LHD}$ temperature profile in the
[OI] 630 nm line-forming layers around log $\tau\sim -1$. 
The differences between our assessment of the impact of granulation on 
the abundances derived from the [OI] 630 nm line in giants, and that of
\citet{collet}, will be further investigated in the future. This might
be rooted in the different binning schemes adopted by the two codes
for the opacity (4 opacity bins for the Stein \& Nordlund code and 6
opacity bins for \cobold). In any case we believe that, at present, it
is safer not to apply any 3D correction to the O abundances derived
for giants from the [OI] 630 nm line. 

In Paper V we found that the mean value of ${\rm [O/Fe]}$ in the
range $-3.6 < {\rm [Fe/H]} < -2.5$ was about $0.72 \pm 0.17$. The
weighted average value in CS22876--032, ${\rm [O/Fe]}=0.83
\pm 0.15$, agrees with this determination. This 
mean value does not exclude a slight increase of the ratio [O/Fe] in
the range $-3.6 < {\rm [Fe/H]} < -2.5$ as seen in the theoretical
predictions of the chemical evolution models presented by
\citet{fra04}. A more detailed discussion of the general behaviour of
oxygen with metallicity will be treated in forthcoming investigations, 
hopefully after 3D corrections will be determined for all the stars, 
and for different lines of oxygen, \ion{Fe}{i} and \ion{Fe}{ii}.

\subsection{The Odd-$Z$ light elements}

The LTE [Na/Fe] abundance ratios of both stars appear consistent
with the Galactic trend of this element in EMP giants 
\citep{cay04}, subgiants and turn-off stars of similar metallicity
\citep{and07}. NLTE corrections are expected to be larger for the
metal-poor giants than for dwarfs. We have estimated the NLTE
corrections\footnote{$\Delta_\mathrm{NLTE}=\log 
\epsilon(\mathrm{X})_\mathrm{NLTE}-\log 
\epsilon(\mathrm{X})_\mathrm{LTE}$} for Na to be
$\Delta_\mathrm{NLTE}\la-0.06$ dex, according to the NLTE corrections 
reported in Table~2 of \citet{and07}. After applying these
corrections, Na abundances of the \mbox{CS 22876--032} dwarfs remain
compatible with those of metal-poor giants and dwarfs, which exhibit
almost a constant ratio [Na/Fe]$\sim -0.20$ in the metallicity range
$-4.0 < {\rm [Fe/H]} < -2.5$.

The LTE [Al/Fe] abundance ratios of the dwarfs in this system are
$\sim 0.5$ dex larger than those in metal-poor giants with
similar iron content. However, aluminium is also expected to exhibit
significant NLTE corrections \citep{bag97}, which might explain this
difference, as was the case for Na. This element also exhibits
an almost constant ratio [Al/Fe]$\sim -0.10$ for giants in the
metallicity range $-4.0 < {\rm [Fe/H]} < -2.5$ when a fixed NLTE
correction of +0.65 dex is considered.

Within the errors, [Sc/Fe] is consistent with, although 0.2--0.3 dex lower 
than, the [Sc/Fe] ratio in EMP giants which shows an almost constant 
ratio [Sc/Fe]$\sim0$. Note that the [X/Fe] ratios given in Table~\ref{tbl7} 
were computed relative to \ion{Fe}{ii} for ionised species. As noted 
in \S~\ref{met}, the \ion{Fe}{ii} abundances are less reliable than
those for \ion{Fe}{i} because the \ion{Fe}{ii} lines in EMP dwarfs are 
very weak, especially for star B in CS 22876--032. Therefore, [X/Fe] 
for ionised species in star B should be regarded with caution. 

\subsection{The $\alpha$ elements}

The [Mg/Fe] ratios in both stars of CS 22876--032 seems to be 
fairly consistent with those found in EMP giants
\citep{cay04}, subgiants and dwarfs (\citet{coh04}, Bonifacio et al.
2007, in prep.), at the level of $\sim0.3-0.4$ dex. 

[Ca/Fe] was derived from both \ion{Ca}{i} and \ion{Ca}{ii} lines, which 
yield similar [Ca/H] abundances, at least for
the primary star. However, although the [Ca/Fe] ratios differ by $\sim
0.2$ dex, they seem to be slightly lower than those measured in
metal-poor giants and dwarfs, where a constant [Ca/Fe]$\sim0.4$ is
seen. 

[Si/Fe] also seems to be low, at [Si/Fe]$\sim -0.1$ for stars A and 
$\sim +0.1$ for the secondary, compared to the constant
[Si/Fe]$\sim 0.4$ for metal-poor giants. However, other metal-poor
dwarfs show similar abundances, which might be related to the
different Si lines used in giants and dwarfs (Bonifacio et al.
2007, in prep.).

Finally, [Ti/Fe] is constant at $\sim 0.3$ in EMP giants. While we find
a similar result for star A, [Ti/Fe] in star B is completely different 
due to its  high \ion{Fe}{ii} abundance. We note that the standard deviation
in [Ti/Fe] is relatively high in both stars ($\sigma\sim 0.3$), although 
the mean abundances are better determined when averaging the results from 
the 19 lines in star A and 12 in B.

\subsection{The iron-peak elements}

Chromium was derived from both \ion{Cr}{i} and \ion{Cr}{ii} lines, and 
we find a difference of $\sim 0.2$ dex for the [Cr/Fe] ratios.
[\ion{Cr}{i}/\ion{Fe}{i}] in CS 22876--032 appears to agree with that 
in other EMP dwarfs, but is slightly higher than seen in EMP giants.

The ratio [Mn/Fe] is consistent with the other EMP giants and dwarfs, at 
[Mn/Fe]$\sim -0.5$. [Ni/Fe] also agrees with the values reported for 
EMP giants, which show a constant [Ni/Fe]$\sim 0$.

Cobalt is found to be slightly enhanced in CS 22876--032 relative to
the gradually increasing trend of [Co/Fe] with decreasing [Fe/H] 
observed in giants, although marginally compatible within the errors.

\subsection{Could CS 22876--032 be a triple system ?}

One might ask whether a third star in CS 22876--032 might contribute 
significantly to the total light and the veiling of the lines of the two 
main components we have discussed so far. Such a star would need to have 
a mass above $\sim$0.5 \Msun in order to have any significant effect on the
observed spectrum. The presence of such a third star can be ruled out by  
two independent pieces of evidence.

First, we have individual spectra of \mbox{CS 22876--032} with S/N ratios
$\sim 100$ in the region of the \ion{Mg}{i} b triplet. These are among
the strongest stellar lines seen in these EMP stars and would be 
at least as strong in the third star. The Cayrel formula predicts that 
any line with an EW above $\sim0.15$ pm would be detected at the $3\sigma$
level. Assuming teff = 5500K and $\log g=4.7$ (from the isochrone) plus 
[Fe/H]= --3.6 and [Mg/Fe]= +0.2 for the hypothetical third star, we need 
to dilute the strongest line of the \ion{Mg}{i} b triplet at 518.36 nm 
which would show an intrinsic EW of $\sim 15.0$ pm. In Appendix \ref{appen_veil}
we define the veiling factors for a triple system and conclude that the
non-detection of the 518.36 nm line requires $f_3 > 100$. We can therefore
conclude that any third star contributes negligibly to the total light of CS
22876--032. 

A second line of evidence is available from the radial velocities, which 
are accurate to $\sim$1 \kms and cover a period of 16 years. Any third star 
of mass comparable to A and B should leave significant trends in the 
velocity residuals from the orbital solution for periods shorter than 
several decades. We have therefore attempted sinusoidal fits to the velocity
residuals for both stars and find periods of the order of 1300 days in both cases.
First, the standard deviations around these fits are 0.83 and 0.51 
\kmso, well below the purely observational errors, which shows that these 
results cannot be statistically significant. Second, a period ratio of 
$\sim3$ between the outer and inner orbits is far too small for a triple 
system to be dynamically stable. A white dwarf in an orbit of much longer period
is a possibility, but would not be detectable in our spectra. 

In summary, we conclude that the abundance results reported here cannot be 
significantly affected by light from a third star in the system -- certainly 
not the discrepant Li abundances of the two stars.

\section{Summary}

Our high-resolution VLT/UVES observations of the double-lined spectroscopic 
binary \mbox{CS 22876--032} confirm that it harbours the most metal-poor 
dwarfs known so far. Our improved orbital elements, together with published 
photometry and theoretical isochrones, enable us to determine stellar
parameters of \teffo$_{,A}=6500\pm100$ K and \loggo$_{A}=4.4\pm0.1$ for the
primary (star A) and \teffo$_{,B}=5900\pm150$ K and \loggo$_{B}=4.6\pm0.1$ for the secondary (star B).

Using 1D OSMARCS models and the {\scshape turbospectrum} code, we 
determine abundances of Li, O, Na, Mg, Al, Si, Ca, Sc, Ti, Cr, Mn, Fe, Co, 
and Ni, correcting the observed spectra for the veiling from the continuum 
flux of the other star. We find [Fe/H]$= -3.66\pm0.16$ and 
[Fe/H]$=-3.57\pm0.25$ for star A and B, respectively. Using CO$^5$BOLD 
model atmospheres to estimate 3D abundance corrections, we compute 
full 3D spectrum synthesis using the {\linfor} code for Li and O to estimate 
the {\ftac} corrections, while we use a horizontal and temporal average of
the 3D model to compute $\langle$3D$\rangle$ abundances with {\scshape turbospectrum} for the rest of elements. In general, we find {\tact} 
corrections to be $\la 0.1$ dex, negative for neutral species and positive 
for ionised species; for Fe in particular, we find corrections to be 
$\sim-0.12$ (A)and $\sim-0.07$ (B). 

The [$\alpha$/Fe] ratios are consistent with our earlier results for 
EMP giants \citep[First Stars V]{cay04}, although Ca and Si are slightly 
low ([X/Fe]$\la 0$), but actually consistent with our results for other EMP dwarfs 
(Bonifacio et al. 2007, in prep.). [Na/Fe] appears consistent with 
both EMP giants and dwarfs when NLTE corrections are considered. The
LTE value of [Al/Fe] is not consistent with those in EMP giants, but Al 
is severely affected by NLTE effects, which may solve this discrepancy. The 
iron-peak elements follow the established trends in EMP giants and dwarfs.

Our high-quality spectra allowed us to measure the Li doublet in both stars 
of CS 22876--032 for the first time. We find NLTE Li abundances of 
$2.18\pm0.05$ and $1.77\pm0.09$ for stars A and B, respectively. While the
Li abundance of star A corresponds to the level of the Spite plateau, the
secondary star has a significantly lower abundance. This discrepancy
may be resolved by assuming that the secondary star has been subject
to significant Li depletion, which, according to standard Li depletion
isochrones, would have been the case if the star were 350K cooler than
assumed by our analysis. Full 3D corrections for Li are estimated to be 
$\sim -0.3$ (A) and $\sim -0.2$ (B); however, these computations were 
performed in LTE, and 3D NLTE corrections are needed to confirm the sign 
and value of these corrections. 

The near-UV part of our VLT/UVES spectra enabled us to measure oxygen 
abundances from the OH bands. We find 1D [O/Fe] values of $2.14\pm0.15$ 
(A) and $1.81\pm0.33$ (B) and compute full 3D corrections for the OH lines, 
which turn out to be --1.5 (A) and --1.0 dex (B). Using these
corrections and the $\langle$3D$\rangle$ \ion{Fe}{i} abundances, we
determine 3D [O/Fe] ratios of $0.77\pm0.15$ (A) and $0.97\pm0.33$ (B). 
These 3D [O/Fe] ratios are consistent with those derived from the [OI] 
line in EMP giants of similar metallicity, where 3D corrections should not
be significant.   

\begin{acknowledgements}

We are grateful to A. Chieffi and M. Limongi, for computing, at our
request, isochrones appropriate for the metallicity of CS 22876--032
and for many interesting discussions on the evolution of extremely low
metallicity dwarfs. J. I., P. B. and H.-G. L. acknowledge support
from the EU contract MEXT-CT-2004-014265 (CIFIST). PB also
acknowledges support from MIUR - PRIN grant 2004025729. BN and JA
thank the Carlsberg Foundation and the Danish Natural Science Research
Council for support for this work. TCB and TS acknowledge partial
support from the US National Science Foundation under grants AST
04-06784, AST 07-07776, as well as from grant PHY 02-15783; Physics
Frontier Center / Joint Institute for Nuclear Astrophysics (JINA).
BN and JA acknowledge support from the Carlsberg Foundation and the 
Danish Natural Science Research Council.
This publication makes use of data products from the Two Micron All
Sky Survey, which is a joint project of the University of
Massachusetts and the Infrared Processing and Analysis
Center/California Institute of Technology, funded by the National
Aeronautics Ans Space Administration and the National Science
Foundation. This work has also made use of the IRAF facilities and the
SIMBAD database, operated at CDS, Strasbourg, France.
      
\end{acknowledgements}

\appendix

\section{Veiling factors for a triple system.\label{appen_veil}}

In this appendix we derive the expression for the veiling factors
of a triple system. In fact they are a trivial extension of 
those for a double system, however since they are not readily
found in any paper or book we know of, we provide them here
for the reader's convienience.

We shall use the following notation:

\begin{tabular}{ll}
$\lambda$ & wavelength\\
$s_i (i=1,2,3) $ & the flux spectrum $s_\lambda$ of\\ 
 &  the $i_{th}$ component\\
$c_i (i=1,2,3) $ & the continuum flux spectrum $c_\lambda$ of \\
 &  the $i_{th}$ component\\
$d_i (i=1,2,4)$ & the line depression = $c_i - s_i$\\
$EW_i (i=1,2,3) $ & the intrinsic equivalent width of a \\ 
 & spectral line of the $i_{th}$ component\\
$EW_i^{obs} (i=1,2,3) $ & the observed equivalent width of a \\  
 & spectral line of the $i_{th}$ component\\
$EW_{123}^{obs}  $ & the observed equivalent width of the \\
 & three components.
\end{tabular}

In any orbital phase we have:

$$EW_{123}^{obs} = \int_{\lambda 1}^{\lambda 2}{d_1+d_2+d_3\over c_1+c_2+c_3} d\lambda$$

Where the interval $[{\lambda 1},{\lambda 2}]$ includes the
desired spectral feauture, which we assume to be isolated.
Considering that the continuum fluxes can be assumed
to be constant over the integration interval
this can be re-written as:\\

$EW_{123}^{obs} = {c_1\over c_1+c_2+c_3} \int_{\lambda 1}^{\lambda
2}{d_1\over c_1} d\lambda + {c_2\over c_1+c_2+c_3} \int_{\lambda
1}^{\lambda 2}{d_2\over c_2} d\lambda +$ 
\vspace{0.01cm}\\

${c_3\over c_1+c_2+c_3} \int_{\lambda 1}^{\lambda 2}{d_3\over c_3}
d\lambda $
\vspace{0.1cm}\\

We now define the veiling factors as
$f_i =  {(c_1+c_2+c_3)/c_i}$ and consider a phase in which
the radial velocities are such that the line of of each 
component is isolated and not blended with the others, then one has:

$$ EW_{123}^{obs} =  EW_{1}^{obs} + EW_{2}^{obs} + EW_{3}^{obs} $$ 

and noting that :
$$\int_{\lambda 1}^{\lambda 2}{d_i\over c_i} d\lambda = EW_i  $$

one has that :

$$  EW_{1}^{obs} + EW_{2}^{obs} + EW_{3}^{obs} = 
{EW_1/f_1}  + 
{EW_2/f_2}  + 
{EW_3/f_3}   
$$

finally considering that in this phase the lines
are not overlapping this equality implies:

$$ EW_i = f_i EW_i^{obs} $$

\Online

\longtab{1}{
\begin{longtable}{llllrrrrrr} 
\caption[]{Line data, equivalent widths, veiling factors and 1D abundances of CS 22876-032A,B}\\ 
\noalign{\smallskip}
\noalign{\smallskip}
\noalign{\smallskip}
\hline
\hline
\noalign{\smallskip}
Specie & $\lambda$  & $\chi$  & $\log gf$ & EW$_{\rm A,obs}$ & $f_{\rm A,1D}$ & A(X)$_{\rm A,1D}$ & EW$_{\rm B,obs}$ & $f_{\rm B,1D}$ & A(X)$_{\rm B,1D}$\\
  & (nm) & (eV) &  & (pm) &  & (dex)  & (pm) &   & (dex) \\
\noalign{\smallskip}
\hline
\hline
\noalign{\smallskip}
\noalign{\smallskip}
\endfirsthead
\caption{Continued.}\\ 
\noalign{\smallskip}
\noalign{\smallskip}
\noalign{\smallskip}
\hline
\hline
\noalign{\smallskip}
Specie & $\lambda$  & $\chi$  & $\log gf$ & EW$_{\rm A,obs}$ & $f_{A,\rm 1D}$ & A(X)$_{\rm A,1D}$ & EW$_{\rm B,obs}$ & $f_{B,\rm 1D}$ & A(X)$_{\rm B,1D}$\\
  & (nm) & (eV) &  & (pm) &  & (dex)  & (pm) &   & (dex) \\
\noalign{\smallskip}
\hline
\hline
\noalign{\smallskip}

\noalign{\smallskip}
\endhead
\noalign{\smallskip}
\hline
\hline
\noalign{\smallskip}
\noalign{\smallskip}
\noalign{\smallskip}
\endfoot
\ion{Na}{i}  & 588.9951 & 0.000 &  0.112 & 1.40 & 1.35 &  2.63 &  --  &  --  &  --   \\ 
\ion{Na}{i}  & 589.5924 & 0.000 & -0.191 & 0.79 & 1.35 &  2.64 & 0.49 & 3.87 &  2.54 \\ 
             &          &       &        &      &      &       &      &      &       \\
\ion{Mg}{i}  & 333.6674 & 2.720 & -1.230 & 1.07 & 1.32 &  4.57 & 0.90 & 4.16 &  4.68 \\ 
\ion{Mg}{i}  & 382.9355 & 2.710 & -0.207 & 5.01 & 1.31 &  4.41 & 2.06 & 4.24 &  4.29 \\ 
\ion{Mg}{i}  & 383.2304 & 2.710 &  0.146 & 6.03 & 1.31 &  4.29 & 2.87 & 4.25 &  4.35 \\ 
\ion{Mg}{i}  & 383.8290 & 2.720 &  0.415 & 6.89 & 1.30 &  4.21 & 2.77 & 4.31 &  4.06 \\ 
\ion{Mg}{i}  & 416.7271 & 4.340 & -1.000 & 0.23 & 1.29 &  4.84 & 0.11 & 4.41 &  4.86 \\ 
\ion{Mg}{i}  & 517.2684 & 2.710 & -0.380 & 4.88 & 1.33 &  4.47 & 2.42 & 4.04 &  4.41 \\ 
\ion{Mg}{i}  & 518.3604 & 2.720 & -0.158 & 5.73 & 1.33 &  4.43 & 3.24 & 4.03 &  4.52 \\ 
\ion{Mg}{i}  & 552.8405 & 4.340 & -0.341 & 0.60 & 1.34 &  4.58 & 0.16 & 3.94 &  4.25 \\ 
\ion{Mg}{i}  & 880.6756 & 4.340 & -0.137 & 0.81 & 1.39 &  4.41 & 0.62 & 3.57 &  4.50 \\ 
             &          &       &        &      &      &       &      &      &       \\
\ion{Al}{i}  & 394.4006 & 0.000 & -0.623 & 1.51 & 1.28 &  2.74 & 1.12 & 4.54 &  2.88 \\ 
\ion{Al}{i}  & 396.1520 & 0.010 & -0.323 & 1.41 & 1.31 &  2.42 & 1.16 & 4.24 &  2.56 \\ 
             &          &       &        &      &      &       &      &      &       \\
\ion{Si}{i}  & 390.5523 & 1.910 & -1.090 & 1.55 & 1.29 &  3.80 & 1.32 & 4.42 &  4.07 \\ 
             &          &       &        &      &      &       &      &      &       \\
\ion{Ca}{i}  & 422.6728 & 0.000 &  0.240 & 3.60 & 1.30 &  2.71 & 1.70 & 4.39 &  2.68 \\ 
             &          &       &        &      &      &       &      &      &       \\
\ion{Ca}{ii} & 317.9331 & 3.150 &  0.512 & 4.08 & 1.30 &  2.62 & 1.36 & 4.28 &  2.57 \\ 
\ion{Ca}{ii} & 370.6024 & 3.120 & -0.480 & 1.41 & 1.34 &  2.58 &  --  &  --  &  --   \\ 
\ion{Ca}{ii} & 373.6902 & 3.150 & -0.173 & 3.22 & 1.33 &  2.84 & 0.68 & 4.04 &  2.53 \\ 
             &          &       &        &      &      &       &      &      &       \\
\ion{Sc}{ii} & 361.3829 & 0.020 &  0.416 & 0.42 & 1.33 & -0.69 & 0.42 & 4.01 & -0.40 \\ 
\ion{Sc}{ii} & 424.6822 & 0.310 &  0.240 & 0.37 & 1.30 & -0.44 & 0.16 & 4.38 & -0.41 \\ 
             &          &       &        &      &      &       &      &      &       \\
\ion{Ti}{ii} & 316.8532 & 0.150 & -0.310 & 2.35 & 1.30 &  1.66 & 0.70 & 4.29 &  1.34 \\ 
\ion{Ti}{ii} & 323.4520 & 0.050 &  0.426 & 4.55 & 1.31 &  1.54 & 1.64 & 4.25 &  1.34 \\ 
\ion{Ti}{ii} & 323.6578 & 0.030 &  0.234 & 4.11 & 1.31 &  1.56 & 1.72 & 4.25 &  1.59 \\ 
\ion{Ti}{ii} & 324.1994 & 0.000 & -0.045 & 3.20 & 1.31 &  1.51 & 1.75 & 4.24 &  1.86 \\ 
\ion{Ti}{ii} & 325.1918 & 0.010 & -0.579 & 1.54 & 1.31 &  1.52 & 0.65 & 4.23 &  1.39 \\ 
\ion{Ti}{ii} & 332.2941 & 0.150 & -0.093 & 2.38 & 1.32 &  1.44 & 1.23 & 4.17 &  1.53 \\ 
\ion{Ti}{ii} & 332.9453 & 0.140 & -0.274 & 2.53 & 1.32 &  1.65 & 1.33 & 4.17 &  1.78 \\ 
\ion{Ti}{ii} & 338.0279 & 0.050 & -0.570 & 3.38 & 1.32 &  2.12 & 1.33 & 4.14 &  1.98 \\ 
\ion{Ti}{ii} & 338.7846 & 0.030 & -0.432 & 2.29 & 1.32 &  1.63 & 1.01 & 4.14 &  1.54 \\ 
\ion{Ti}{ii} & 344.4314 & 0.150 & -0.810 & 1.02 & 1.32 &  1.63 &  --  &  --  &  --   \\ 
\ion{Ti}{ii} & 345.6388 & 2.060 & -0.230 & 0.41 & 1.32 &  2.32 &  --  &  --  &  --   \\ 
\ion{Ti}{ii} & 347.7187 & 0.120 & -0.967 & 1.17 & 1.32 &  1.83 & 0.58 & 4.08 &  1.76 \\ 
\ion{Ti}{ii} & 348.9741 & 0.140 & -1.920 & 0.40 & 1.33 &  2.27 &  --  &  --  &  --   \\ 
\ion{Ti}{ii} & 350.0340 & 0.120 & -2.020 & 0.37 & 1.33 &  2.32 &  --  &  --  &  --   \\ 
\ion{Ti}{ii} & 375.9296 & 0.610 &  0.270 & 2.39 & 1.32 &  1.33 & 0.94 & 4.16 &  1.26 \\ 
\ion{Ti}{ii} & 376.1323 & 0.570 &  0.170 & 2.36 & 1.32 &  1.39 & 1.07 & 4.16 &  1.43 \\ 
\ion{Ti}{ii} & 391.3468 & 1.120 & -0.410 & 0.39 & 1.28 &  1.48 &  --  &  --  &  --   \\ 
\ion{Ti}{ii} & 402.8343 & 1.890 & -0.990 & 0.15 & 1.29 &  2.30 &  --  &  --  &  --   \\ 
\ion{Ti}{ii} & 518.8680 & 1.580 & -1.050 & 0.14 & 1.33 &  2.01 &  --  &  --  &  --   \\ 
             &          &       &        &      &      &       &      &      &       \\
\ion{Cr}{i}  & 357.8684 & 0.000 &  0.409 & 0.85 & 1.33 &  1.79 & 0.61 & 4.03 &  1.61 \\ 
\ion{Cr}{i}  & 425.4332 & 0.000 & -0.110 & 0.45 & 1.30 &  1.83 & 0.28 & 4.37 &  1.63 \\ 
\ion{Cr}{i}  & 427.4796 & 0.000 & -0.230 & 0.32 & 1.30 &  1.79 & 0.16 & 4.36 &  1.48 \\ 
\ion{Cr}{i}  & 428.9716 & 0.000 & -0.360 & 0.46 & 1.30 &  2.09 & 0.58 & 4.35 &  2.27 \\ 
\ion{Cr}{i}  & 520.6038 & 0.940 &  0.020 & 0.13 & 1.33 &  1.92 &  --  &  --  &  --   \\ 
             &          &       &        &      &      &       &      &      &       \\
\ion{Cr}{ii} & 313.2053 & 2.480 &  0.451 & 2.85 & 1.30 &  2.29 & 1.10 & 4.32 &  2.37 \\ 
             &          &       &        &      &      &       &      &      &       \\
\ion{Mn}{ii} & 344.1988 & 1.780 & -0.270 & 0.54 & 1.32 &  1.38 & 0.32 & 4.11 &  1.53 \\ 
             &          &       &        &      &      &       &      &      &       \\
\ion{Fe}{i}  & 347.5450 & 0.090 & -1.054 & 3.24 & 1.32 &  3.99 & 1.89 & 4.09 &  4.07 \\ 
\ion{Fe}{i}  & 347.6702 & 0.120 & -1.507 & 1.96 & 1.32 &  4.05 & 1.40 & 4.08 &  4.07 \\ 
\ion{Fe}{i}  & 349.0574 & 0.050 & -1.105 & 2.91 & 1.33 &  3.90 & 1.93 & 4.07 &  4.11 \\ 
\ion{Fe}{i}  & 356.5379 & 0.960 & -0.133 & 2.96 & 1.33 &  3.75 & 1.90 & 4.04 &  3.92 \\ 
\ion{Fe}{i}  & 358.1193 & 0.860 &  0.406 & 4.17 & 1.33 &  3.55 & 2.81 & 4.03 &  3.89 \\ 
\ion{Fe}{i}  & 360.8859 & 1.010 & -0.100 & 3.39 & 1.33 &  3.91 & 2.17 & 4.01 &  4.13 \\ 
\ion{Fe}{i}  & 361.8768 & 0.990 & -0.003 & 3.55 & 1.33 &  3.85 & 2.25 & 4.00 &  4.06 \\ 
\ion{Fe}{i}  & 364.7843 & 0.920 & -0.194 & 3.29 & 1.33 &  3.72 & 2.24 & 4.00 &  4.09 \\ 
\ion{Fe}{i}  & 381.5840 & 1.490 &  0.237 & 3.79 & 1.30 &  3.88 & 1.61 & 4.34 &  3.77 \\ 
\ion{Fe}{i}  & 382.0425 & 0.860 &  0.119 & 5.15 & 1.30 &  3.89 & 2.00 & 4.30 &  3.63 \\ 
\ion{Fe}{i}  & 382.5881 & 0.920 & -0.037 & 4.10 & 1.31 &  3.76 & 2.00 & 4.26 &  3.82 \\ 
\ion{Fe}{i}  & 382.7823 & 1.560 &  0.062 & 2.57 & 1.31 &  3.79 & 1.29 & 4.25 &  3.71 \\ 
\ion{Fe}{i}  & 384.0438 & 0.990 & -0.506 & 2.26 & 1.30 &  3.76 & 1.11 & 4.34 &  3.60 \\ 
\ion{Fe}{i}  & 384.1048 & 1.610 & -0.045 & 2.05 & 1.30 &  3.78 & 0.88 & 4.35 &  3.54 \\ 
\ion{Fe}{i}  & 384.9967 & 1.010 & -0.871 & 1.26 & 1.29 &  3.79 & 1.09 & 4.46 &  3.99 \\ 
\ion{Fe}{i}  & 385.6372 & 0.050 & -1.286 & 3.23 & 1.28 &  3.96 & 1.76 & 4.52 &  4.16 \\ 
\ion{Fe}{i}  & 385.9911 & 0.000 & -0.710 & 5.44 & 1.28 &  4.03 & 2.42 & 4.54 &  4.08 \\ 
\ion{Fe}{i}  & 386.5523 & 1.010 & -0.982 & 1.03 & 1.28 &  3.79 &  --  &  --  &  --   \\ 
\ion{Fe}{i}  & 387.2501 & 0.990 & -0.928 & 1.58 & 1.29 &  3.95 & 0.99 & 4.49 &  3.94 \\ 
\ion{Fe}{i}  & 387.8018 & 0.960 & -0.914 & 1.59 & 1.30 &  3.91 & 1.29 & 4.36 &  4.13 \\ 
\ion{Fe}{i}  & 388.6282 & 0.050 & -1.076 & 2.64 & 1.32 &  3.60 & 2.18 & 4.17 &  4.17 \\ 
\ion{Fe}{i}  & 388.7048 & 0.920 & -1.144 & 1.02 & 1.32 &  3.87 & 1.15 & 4.16 &  4.15 \\ 
\ion{Fe}{i}  & 389.5656 & 0.110 & -1.670 & 1.59 & 1.31 &  3.92 & 1.43 & 4.19 &  4.15 \\ 
\ion{Fe}{i}  & 389.9707 & 0.090 & -1.531 & 2.16 & 1.31 &  3.94 & 1.66 & 4.28 &  4.23 \\ 
\ion{Fe}{i}  & 390.2946 & 1.560 & -0.466 & 1.10 & 1.30 &  3.79 & 0.91 & 4.35 &  3.93 \\ 
\ion{Fe}{i}  & 390.6480 & 0.110 & -2.243 & 0.77 & 1.29 &  4.09 & 0.63 & 4.44 &  4.05 \\ 
\ion{Fe}{i}  & 392.0258 & 0.120 & -1.746 & 1.39 & 1.28 &  3.91 & 1.28 & 4.58 &  4.20 \\ 
\ion{Fe}{i}  & 392.7920 & 0.110 & -1.522 & 2.26 & 1.28 &  3.97 & 1.44 & 4.58 &  4.12 \\ 
\ion{Fe}{i}  & 404.5812 & 1.490 &  0.280 & 3.96 & 1.29 &  3.85 & 1.38 & 4.47 &  3.52 \\ 
\ion{Fe}{i}  & 406.3594 & 1.560 &  0.062 & 2.88 & 1.29 &  3.84 & 1.49 & 4.47 &  3.90 \\ 
\ion{Fe}{i}  & 407.1738 & 1.610 & -0.022 & 2.54 & 1.29 &  3.87 & 1.26 & 4.44 &  3.83 \\ 
\ion{Fe}{i}  & 413.2058 & 1.610 & -0.675 & 0.98 & 1.29 &  3.96 & 0.67 & 4.42 &  3.96 \\ 
\ion{Fe}{i}  & 414.3868 & 1.560 & -0.511 & 1.24 & 1.29 &  3.87 & 1.07 & 4.43 &  4.09 \\ 
\ion{Fe}{i}  & 420.2029 & 1.490 & -0.708 & 0.93 & 1.29 &  3.85 & 0.70 & 4.40 &  3.89 \\ 
\ion{Fe}{i}  & 425.0787 & 1.560 & -0.714 & 0.58 & 1.30 &  3.69 & 0.49 & 4.37 &  3.74 \\ 
\ion{Fe}{i}  & 426.0474 & 2.400 &  0.109 & 0.72 & 1.30 &  3.71 & 0.28 & 4.37 &  3.43 \\ 
\ion{Fe}{i}  & 427.1761 & 1.490 & -0.164 & 2.43 & 1.30 &  3.86 & 1.23 & 4.36 &  3.78 \\ 
\ion{Fe}{i}  & 432.5762 & 1.610 &  0.006 & 2.19 & 1.31 &  3.73 & 1.39 & 4.19 &  3.81 \\ 
             &          &       &        &      &      &       &      &      &       \\
\ion{Fe}{ii} & 318.6738 & 1.700 & -1.710 & 1.69 & 1.30 &  4.13 & 0.90 & 4.28 &  4.37 \\ 
\ion{Fe}{ii} & 319.2909 & 1.670 & -1.950 & 1.37 & 1.31 &  4.20 & 0.29 & 4.27 &  3.87 \\ 
\ion{Fe}{ii} & 319.3799 & 1.720 & -1.720 & 1.30 & 1.31 &  4.01 & 0.70 & 4.27 &  4.21 \\ 
\ion{Fe}{ii} & 319.6070 & 1.670 & -1.660 & 1.61 & 1.31 &  4.03 & 0.44 & 4.27 &  3.80 \\ 
\ion{Fe}{ii} & 321.0444 & 1.720 & -1.790 & 2.05 & 1.31 &  4.36 & 0.95 & 4.26 &  4.52 \\ 
\ion{Fe}{ii} & 321.3309 & 1.700 & -1.230 & 1.62 & 1.31 &  3.62 &  --  &  --  &  --   \\ 
\ion{Fe}{ii} & 322.7742 & 1.670 & -1.130 & 3.29 & 1.31 &  4.05 & 1.94 & 4.26 &  4.81 \\ 
\ion{Fe}{ii} & 325.5887 & 0.990 & -2.500 & 1.47 & 1.31 &  4.19 & 0.76 & 4.22 &  4.31 \\ 
\ion{Fe}{ii} & 327.7348 & 0.990 & -2.470 & 1.33 & 1.31 &  4.11 & 0.72 & 4.20 &  4.23 \\ 
\ion{Fe}{ii} & 423.3172 & 2.580 & -1.900 & 0.51 & 1.30 &  4.26 &  --  &  --  &  --   \\ 
\ion{Fe}{ii} & 492.3927 & 2.890 & -1.320 & 0.36 & 1.32 &  3.77 &  --  &  --  &  --   \\ 
\ion{Fe}{ii} & 501.8440 & 2.890 & -1.220 & 0.42 & 1.33 &  3.73 & 0.27 & 4.08 &  4.05 \\ 
\ion{Fe}{ii} & 516.9033 & 2.890 & -0.870 & 0.65 & 1.33 &  3.58 &  --  &  --  &  --   \\ 
             &          &       &        &      &      &       &      &      &       \\
\ion{Co}{i}  & 340.5114 & 0.430 &  0.250 & 1.17 & 1.32 &  1.93 & 2.44 & 4.12 &  2.10 \\ 
\ion{Co}{i}  & 341.2333 & 0.510 &  0.030 & 1.09 & 1.32 &  2.05 & 0.86 & 4.12 &  2.02 \\ 
\ion{Co}{i}  & 345.3508 & 0.430 &  0.380 & 1.54 & 1.32 &  1.99 & 1.41 & 4.11 &  1.95 \\ 
\ion{Co}{i}  & 349.5681 & 0.630 & -0.270 & 0.65 & 1.33 &  2.11 &  --  &  --  &  --   \\ 
\ion{Co}{i}  & 350.2278 & 0.430 &  0.070 & 1.40 & 1.33 &  2.11 &  --  &  --  &  --   \\ 
\ion{Co}{i}  & 399.5302 & 0.920 & -0.220 & 0.24 & 1.29 &  2.00 & 0.19 & 4.48 &  1.94 \\ 
\ion{Co}{i}  & 412.1311 & 0.920 & -0.320 & 0.18 & 1.31 &  1.90 &  --  &  --  &  --   \\ 
             &          &       &        &      &      &       &      &      &       \\
\ion{Ni}{i}  & 339.2983 & 0.030 & -0.540 & 2.66 & 1.32 &  2.90 & 1.57 & 4.13 &  2.92 \\ 
\ion{Ni}{i}  & 343.3554 & 0.030 & -0.668 & 1.89 & 1.32 &  2.77 & 1.48 & 4.10 &  2.93 \\ 
\ion{Ni}{i}  & 345.2885 & 0.110 & -0.910 & 1.14 & 1.32 &  2.78 & 1.13 & 4.11 &  2.93 \\ 
\ion{Ni}{i}  & 345.8456 & 0.210 & -0.223 & 2.25 & 1.32 &  2.62 & 1.32 & 4.10 &  2.51 \\ 
\ion{Ni}{i}  & 346.1649 & 0.030 & -0.347 & 2.55 & 1.32 &  2.67 & 1.51 & 4.10 &  2.66 \\ 
\ion{Ni}{i}  & 349.2954 & 0.110 & -0.250 & 2.39 & 1.33 &  2.59 & 1.76 & 4.07 &  2.84 \\ 
\ion{Ni}{i}  & 351.5049 & 0.110 & -0.211 & 2.74 & 1.33 &  2.67 & 1.74 & 4.06 &  2.77 \\ 
\ion{Ni}{i}  & 352.4535 & 0.030 &  0.008 & 3.90 & 1.33 &  2.77 & 2.12 & 4.05 &  2.80 \\ 
\ion{Ni}{i}  & 361.0461 & 0.110 & -1.149 & 0.95 & 1.33 &  2.92 & 1.07 & 4.01 &  3.05 \\ 
\ion{Ni}{i}  & 361.9386 & 0.420 &  0.035 & 2.02 & 1.33 &  2.46 & 1.72 & 4.00 &  2.75 \\ 
\ion{Ni}{i}  & 380.7138 & 0.420 & -1.180 & 1.11 & 1.30 &  3.17 & 0.30 & 4.38 &  2.58 \\ 
\ion{Ni}{i}  & 385.8292 & 0.420 & -0.970 & 0.73 & 1.28 &  2.74 & 0.68 & 4.54 &  2.85 \\ 
\label{BIGtable}  
\end{longtable}
}

\begin{table*}
\centering
\caption[]{Hyperfine structure of Sc and Co}
\scriptsize{
\begin{tabular}{lrrrlrrrlrrr}     
\noalign{\smallskip}
\noalign{\smallskip}
\noalign{\smallskip}
\hline
\hline
\noalign{\smallskip}
Specie & $\lambda$ (nm) & $\chi$ (eV) & $\log gf$ & Specie & $\lambda$ (nm) & $\chi$ (eV) & $\log gf$ & Specie & $\lambda$ (nm) & $\chi$ (eV) & $\log gf$ \\
\noalign{\smallskip}
\hline
\hline
\noalign{\smallskip}
\noalign{\smallskip}
\ion{Sc}{ii} & 361.3815 & 0.022 & -0.126 & \ion{Co}{i}  & 341.2328 & 0.514 & -0.171 & \ion{Co}{i}  & 350.2241 & 0.432 & -0.221 \\ 
\ion{Sc}{ii} & 361.3814 & 0.022 & -0.141 & \ion{Co}{i}  & 341.2325 & 0.514 & -0.151 & \ion{Co}{i}  & 350.2263 & 0.432 & -0.110 \\
\ion{Sc}{ii} & 361.3819 & 0.022 & -0.174 & \ion{Co}{i}  & 341.2333 & 0.514 & -0.241 & \ion{Co}{i}  & 350.2255 & 0.432 & -0.132 \\
\ion{Sc}{ii} & 361.3817 & 0.022 & -0.108 & \ion{Co}{i}  & 341.2330 & 0.514 & -0.151 & \ion{Co}{i}  & 350.2245 & 0.432 & -0.231 \\
\ion{Sc}{ii} & 361.3815 & 0.022 & -0.107 & \ion{Co}{i}  & 341.2325 & 0.514 & -0.131 & \ion{Co}{i}  & 350.2272 & 0.432 & -0.095 \\
\ion{Sc}{ii} & 361.3823 & 0.022 & -0.163 & \ion{Co}{i}  & 341.2338 & 0.514 & -0.225 & \ion{Co}{i}  & 350.2262 & 0.432 & -0.133 \\
\ion{Sc}{ii} & 361.3821 & 0.022 & -0.096 & \ion{Co}{i}  & 341.2333 & 0.514 & -0.141 & \ion{Co}{i}  & 350.2251 & 0.432 & -0.251 \\
\ion{Sc}{ii} & 361.3817 & 0.022 & -0.083 & \ion{Co}{i}  & 341.2326 & 0.514 & -0.114 & \ion{Co}{i}  & 350.2283 & 0.432 & -0.082 \\
\ion{Sc}{ii} & 361.3828 & 0.022 & -0.165 & \ion{Co}{i}  & 341.2343 & 0.514 & -0.225 & \ion{Co}{i}  & 350.2271 & 0.432 & -0.140 \\
\ion{Sc}{ii} & 361.3825 & 0.022 & -0.090 & \ion{Co}{i}  & 341.2336 & 0.514 & -0.136 & \ion{Co}{i}  & 350.2257 & 0.432 & -0.291 \\
\ion{Sc}{ii} & 361.3821 & 0.022 & -0.065 & \ion{Co}{i}  & 341.2328 & 0.514 & -0.099 & \ion{Co}{i}  & 350.2295 & 0.432 & -0.071 \\
\ion{Sc}{ii} & 361.3835 & 0.022 & -0.176 & \ion{Co}{i}  & 341.2349 & 0.514 & -0.235 & \ion{Co}{i}  & 350.2281 & 0.432 & -0.161 \\
\ion{Sc}{ii} & 361.3831 & 0.022 & -0.089 & \ion{Co}{i}  & 341.2341 & 0.514 & -0.137 & \ion{Co}{i}  & 350.2308 & 0.432 & -0.060 \\
\ion{Sc}{ii} & 361.3826 & 0.022 & -0.049 & \ion{Co}{i}  & 341.2330 & 0.514 & -0.086 &   	   &	      &       &        \\
\ion{Sc}{ii} & 361.3843 & 0.022 & -0.198 & \ion{Co}{i}  & 341.2356 & 0.514 & -0.256 & \ion{Co}{i}  & 384.5473 & 0.923 & -0.180 \\
\ion{Sc}{ii} & 361.3838 & 0.022 & -0.096 & \ion{Co}{i}  & 341.2346 & 0.514 & -0.144 & \ion{Co}{i}  & 384.5475 & 0.923 & -0.173 \\
\ion{Sc}{ii} & 361.3832 & 0.022 & -0.036 & \ion{Co}{i}  & 341.2334 & 0.514 & -0.075 & \ion{Co}{i}  & 384.5470 & 0.923 & -0.153 \\
\ion{Sc}{ii} & 361.3853 & 0.022 & -0.238 & \ion{Co}{i}  & 341.2364 & 0.514 & -0.295 & \ion{Co}{i}  & 384.5479 & 0.923 & -0.243 \\
\ion{Sc}{ii} & 361.3847 & 0.022 & -0.117 & \ion{Co}{i}  & 341.2352 & 0.514 & -0.165 & \ion{Co}{i}  & 384.5474 & 0.923 & -0.153 \\
\ion{Sc}{ii} & 361.3840 & 0.022 & -0.024 & \ion{Co}{i}  & 341.2338 & 0.514 & -0.064 & \ion{Co}{i}  & 384.5468 & 0.923 & -0.133 \\
             &          &       &        & 		&	   &	   &	    & \ion{Co}{i}  & 384.5480 & 0.923 & -0.227 \\
\ion{Sc}{ii} & 424.6832 & 0.315 & -0.096 & \ion{Co}{i}  & 344.9178 & 0.581 & -0.174 & \ion{Co}{i}  & 384.5474 & 0.923 & -0.143 \\
\ion{Sc}{ii} & 424.6836 & 0.315 & -0.080 & \ion{Co}{i}  & 344.9172 & 0.581 & -0.149 & \ion{Co}{i}  & 384.5465 & 0.923 & -0.116 \\
\ion{Sc}{ii} & 424.6836 & 0.315 & -0.039 & \ion{Co}{i}  & 344.9183 & 0.581 & -0.149 & \ion{Co}{i}  & 384.5481 & 0.923 & -0.227 \\
\ion{Sc}{ii} & 424.6839 & 0.315 & -0.081 & \ion{Co}{i}  & 344.9176 & 0.581 & -0.339 & \ion{Co}{i}  & 384.5473 & 0.923 & -0.138 \\
\ion{Sc}{ii} & 424.6839 & 0.315 & -0.078 & \ion{Co}{i}  & 344.9167 & 0.581 & -0.129 & \ion{Co}{i}  & 384.5462 & 0.923 & -0.101 \\
\ion{Sc}{ii} & 424.6841 & 0.315 & -0.098 & \ion{Co}{i}  & 344.9183 & 0.581 & -0.129 & \ion{Co}{i}  & 384.5482 & 0.923 & -0.237 \\
\ion{Sc}{ii} & 424.6842 & 0.315 & -0.148 & \ion{Co}{i}  & 344.9173 & 0.581 & -0.215 & \ion{Co}{i}  & 384.5471 & 0.923 & -0.139 \\
\ion{Sc}{ii} & 424.6843 & 0.315 & -0.291 & \ion{Co}{i}  & 344.9161 & 0.581 & -0.122 & \ion{Co}{i}  & 384.5458 & 0.923 & -0.088 \\
\ion{Sc}{ii} & 424.6843 & 0.315 & -0.096 & \ion{Co}{i}  & 344.9182 & 0.581 & -0.122 & \ion{Co}{i}  & 384.5483 & 0.923 & -0.257 \\
\ion{Sc}{ii} & 424.6844 & 0.315 & -0.116 & \ion{Co}{i}  & 344.9170 & 0.581 & -0.143 & \ion{Co}{i}  & 384.5470 & 0.923 & -0.146 \\
\ion{Sc}{ii} & 424.6845 & 0.315 & -0.080 & \ion{Co}{i}  & 344.9154 & 0.581 & -0.125 & \ion{Co}{i}  & 384.5455 & 0.923 & -0.077 \\
\ion{Sc}{ii} & 424.6846 & 0.315 & -0.098 & \ion{Co}{i}  & 344.9181 & 0.581 & -0.125 & \ion{Co}{i}  & 384.5483 & 0.923 & -0.297 \\
\ion{Sc}{ii} & 424.6846 & 0.315 & -0.081 & \ion{Co}{i}  & 344.9165 & 0.581 & -0.103 & \ion{Co}{i}  & 384.5468 & 0.923 & -0.167 \\
             &          &       &        & \ion{Co}{i}  & 344.9146 & 0.581 & -0.144 & \ion{Co}{i}  & 384.5450 & 0.923 & -0.066 \\
\ion{Co}{i}  & 340.5074 & 0.432 & -0.139 & \ion{Co}{i}  & 344.9178 & 0.581 & -0.144 &   	   &	      &       &        \\
\ion{Co}{i}  & 340.5072 & 0.432 & -0.159 & \ion{Co}{i}  & 344.9159 & 0.581 & -0.074 & \ion{Co}{i}  & 399.5299 & 0.923 & -0.203 \\
\ion{Co}{i}  & 340.5081 & 0.432 & -0.159 & 		&	   &	   &	    & \ion{Co}{i}  & 399.5304 & 0.923 & -0.203 \\
\ion{Co}{i}  & 340.5079 & 0.432 & -0.138 & \ion{Co}{i}  & 345.3472 & 0.432 & -0.105 & \ion{Co}{i}  & 399.5302 & 0.923 & -0.305 \\
\ion{Co}{i}  & 340.5075 & 0.432 & -0.137 & \ion{Co}{i}  & 345.3479 & 0.432 & -0.150 & \ion{Co}{i}  & 399.5297 & 0.923 & -0.175 \\
\ion{Co}{i}  & 340.5088 & 0.432 & -0.137 & \ion{Co}{i}  & 345.3475 & 0.432 & -0.093 & \ion{Co}{i}  & 399.5306 & 0.923 & -0.175 \\
\ion{Co}{i}  & 340.5085 & 0.432 & -0.124 & \ion{Co}{i}  & 345.3489 & 0.432 & -0.250 & \ion{Co}{i}  & 399.5302 & 0.923 & -0.235 \\
\ion{Co}{i}  & 340.5081 & 0.432 & -0.126 & \ion{Co}{i}  & 345.3485 & 0.432 & -0.129 & \ion{Co}{i}  & 399.5296 & 0.923 & -0.161 \\
\ion{Co}{i}  & 340.5098 & 0.432 & -0.126 & \ion{Co}{i}  & 345.3480 & 0.432 & -0.081 & \ion{Co}{i}  & 399.5308 & 0.923 & -0.161 \\
\ion{Co}{i}  & 340.5093 & 0.432 & -0.107 & \ion{Co}{i}  & 345.3498 & 0.432 & -0.229 & \ion{Co}{i}  & 399.5302 & 0.923 & -0.190 \\
\ion{Co}{i}  & 340.5088 & 0.432 & -0.122 & \ion{Co}{i}  & 345.3493 & 0.432 & -0.119 & \ion{Co}{i}  & 399.5294 & 0.923 & -0.154 \\
\ion{Co}{i}  & 340.5110 & 0.432 & -0.122 & \ion{Co}{i}  & 345.3488 & 0.432 & -0.070 & \ion{Co}{i}  & 399.5310 & 0.923 & -0.154 \\
\ion{Co}{i}  & 340.5104 & 0.432 & -0.090 & \ion{Co}{i}  & 345.3510 & 0.432 & -0.226 & \ion{Co}{i}  & 399.5302 & 0.923 & -0.157 \\
\ion{Co}{i}  & 340.5097 & 0.432 & -0.123 & \ion{Co}{i}  & 345.3504 & 0.432 & -0.115 & \ion{Co}{i}  & 399.5292 & 0.923 & -0.153 \\
\ion{Co}{i}  & 340.5123 & 0.432 & -0.123 & \ion{Co}{i}  & 345.3497 & 0.432 & -0.059 & \ion{Co}{i}  & 399.5313 & 0.923 & -0.153 \\
\ion{Co}{i}  & 340.5116 & 0.432 & -0.074 & \ion{Co}{i}  & 345.3524 & 0.432 & -0.234 & \ion{Co}{i}  & 399.5302 & 0.923 & -0.131 \\
\ion{Co}{i}  & 340.5108 & 0.432 & -0.131 & \ion{Co}{i}  & 345.3517 & 0.432 & -0.117 & \ion{Co}{i}  & 399.5290 & 0.923 & -0.160 \\
\ion{Co}{i}  & 340.5139 & 0.432 & -0.131 & \ion{Co}{i}  & 345.3508 & 0.432 & -0.050 & \ion{Co}{i}  & 399.5315 & 0.923 & -0.160 \\
\ion{Co}{i}  & 340.5131 & 0.432 & -0.060 & \ion{Co}{i}  & 345.3540 & 0.432 & -0.253 & \ion{Co}{i}  & 399.5302 & 0.923 & -0.109 \\
\ion{Co}{i}  & 340.5121 & 0.432 & -0.152 & \ion{Co}{i}  & 345.3531 & 0.432 & -0.125 & \ion{Co}{i}  & 399.5288 & 0.923 & -0.180 \\
\ion{Co}{i}  & 340.5156 & 0.432 & -0.152 & \ion{Co}{i}  & 345.3522 & 0.432 & -0.041 & \ion{Co}{i}  & 399.5317 & 0.923 & -0.180 \\
\ion{Co}{i}  & 340.5147 & 0.432 & -0.046 & \ion{Co}{i}  & 345.3558 & 0.432 & -0.292 & \ion{Co}{i}  & 399.5302 & 0.923 & -0.090 \\
             &          &       &        & \ion{Co}{i}  & 345.3548 & 0.432 & -0.147 &   	   &	      &       &        \\
\ion{Co}{i}  & 340.9155 & 0.514 & -0.204 & \ion{Co}{i}  & 345.3537 & 0.432 & -0.032 & \ion{Co}{i}  & 412.1330 & 0.923 & -0.213 \\
\ion{Co}{i}  & 340.9159 & 0.514 & -0.204 & 		&	   &	   &	    & \ion{Co}{i}  & 412.1332 & 0.923 & -0.206 \\
\ion{Co}{i}  & 340.9157 & 0.514 & -0.306 & \ion{Co}{i}  & 349.5710 & 0.629 & -0.147 & \ion{Co}{i}  & 412.1326 & 0.923 & -0.186 \\
\ion{Co}{i}  & 340.9154 & 0.514 & -0.176 & \ion{Co}{i}  & 349.5703 & 0.629 & -0.147 & \ion{Co}{i}  & 412.1337 & 0.923 & -0.276 \\
\ion{Co}{i}  & 340.9162 & 0.514 & -0.176 & \ion{Co}{i}  & 349.5693 & 0.629 & -0.177 & \ion{Co}{i}  & 412.1331 & 0.923 & -0.186 \\
\ion{Co}{i}  & 340.9159 & 0.514 & -0.236 & \ion{Co}{i}  & 349.5707 & 0.629 & -0.165 & \ion{Co}{i}  & 412.1323 & 0.923 & -0.166 \\
\ion{Co}{i}  & 340.9155 & 0.514 & -0.162 & \ion{Co}{i}  & 349.5697 & 0.629 & -0.131 & \ion{Co}{i}  & 412.1338 & 0.923 & -0.260 \\

\ion{Co}{i}  & 340.9167 & 0.514 & -0.162 & \ion{Co}{i}  & 349.5684 & 0.629 & -0.134 & \ion{Co}{i}  & 412.1329 & 0.923 & -0.176 \\
\ion{Co}{i}  & 340.9163 & 0.514 & -0.191 & \ion{Co}{i}  & 349.5702 & 0.629 & -0.190 & \ion{Co}{i}  & 412.1318 & 0.923 & -0.149 \\
\ion{Co}{i}  & 340.9157 & 0.514 & -0.155 & \ion{Co}{i}  & 349.5689 & 0.629 & -0.130 & \ion{Co}{i}  & 412.1338 & 0.923 & -0.260 \\
\ion{Co}{i}  & 340.9173 & 0.514 & -0.155 & \ion{Co}{i}  & 349.5673 & 0.629 & -0.105 & \ion{Co}{i}  & 412.1327 & 0.923 & -0.171 \\
\ion{Co}{i}  & 340.9167 & 0.514 & -0.158 & \ion{Co}{i}  & 349.5695 & 0.629 & -0.233 & \ion{Co}{i}  & 412.1313 & 0.923 & -0.134 \\
\ion{Co}{i}  & 340.9160 & 0.514 & -0.154 & \ion{Co}{i}  & 349.5679 & 0.629 & -0.146 & \ion{Co}{i}  & 412.1338 & 0.923 & -0.270 \\
\ion{Co}{i}  & 340.9180 & 0.514 & -0.154 & \ion{Co}{i}  & 349.5659 & 0.629 & -0.084 & \ion{Co}{i}  & 412.1324 & 0.923 & -0.172 \\
\ion{Co}{i}  & 340.9173 & 0.514 & -0.132 & 		&	   &	   &	    & \ion{Co}{i}  & 412.1308 & 0.923 & -0.121 \\
\ion{Co}{i}  & 340.9164 & 0.514 & -0.161 & \ion{Co}{i}  & 350.2243 & 0.432 & -0.174 & \ion{Co}{i}  & 412.1338 & 0.923 & -0.290 \\
\ion{Co}{i}  & 340.9188 & 0.514 & -0.161 & \ion{Co}{i}  & 350.2242 & 0.432 & -0.167 & \ion{Co}{i}  & 412.1321 & 0.923 & -0.179 \\
\ion{Co}{i}  & 340.9179 & 0.514 & -0.110 & \ion{Co}{i}  & 350.2238 & 0.432 & -0.237 & \ion{Co}{i}  & 412.1301 & 0.923 & -0.110 \\
\ion{Co}{i}  & 340.9169 & 0.514 & -0.181 & \ion{Co}{i}  & 350.2248 & 0.432 & -0.147 & \ion{Co}{i}  & 412.1336 & 0.923 & -0.330 \\
\ion{Co}{i}  & 340.9197 & 0.514 & -0.181 & \ion{Co}{i}  & 350.2245 & 0.432 & -0.147 & \ion{Co}{i}  & 412.1316 & 0.923 & -0.200 \\
\ion{Co}{i}  & 340.9187 & 0.514 & -0.091 & \ion{Co}{i}  & 350.2239 & 0.432 & -0.221 & \ion{Co}{i}  & 412.1294 & 0.923 & -0.099 \\
             &          &       &        & \ion{Co}{i}  & 350.2255 & 0.432 & -0.127 &   	   &	      &       &        \\
\ion{Co}{i}  & 341.2325 & 0.514 & -0.178 & \ion{Co}{i}  & 350.2249 & 0.432 & -0.137 &   	   &	      &       &        \\
\noalign{\smallskip}
\hline
\hline
\noalign{\smallskip}
\noalign{\smallskip}
\noalign{\smallskip}
\end{tabular}
}
\label{tblhfs}  
\end{table*}


\begin{thebibliography}{}

\bibitem[Alvarez \& Plez (1998)]{aap98} 
Alvarez R., Plez B., 1998, \aap, 330, 1109

\bibitem[Andrievsky et al. (2007)]{and07} 
Andrievsky, S. M., Spite, M., Korotin, S. A, \AA. 2007, \aap, 464,
1081

\bibitem[Asplund \& Garc{\'\i}a P{\'e}rez (2001)]{agp01} 
Asplund, M., \& Garc{\'{\i}}a P{\'e}rez, A.~E.\ 2001, \aap, 372, 601 

\bibitem[Asplund et al. (1997)]{asp97}
Asplund, M., Gustafsson, B., Kiselman, D., Eriksson, K. 1997, \aap,
318, 521  

\bibitem[Asplund et al. (1999)]{asp99} 
Asplund, M., Nordlund, {\AA}., Trampedach, R., Stein, R.F.\ 1999, \aap, 346, L17 

\bibitem[Asplund et al. (2003)]{asp03} 
Asplund, M., Carlsson, M., \& Botnen, A.~V.\ 2003, \aap, 399, L31 

\bibitem[Asplund et al. (2006)]{asp06}
Asplund, M., Lambert, D. L., Nissen, P. E., Primas, F., \& Smith, V. V.
2006, \apj, 644, 229  

\bibitem[Baum\"uller \& Gehren (1997)]{bag97}
Baum\"uller, D., \& Gehren, T. 1997, \aap, 325, 1088

\bibitem[Barklem et al. (2000)]{bar00} 
Barklem, P.~S., Piskunov, N., \& O'Mara, B.~J.\ 2000, \aap, 363, 1091 

\bibitem[Barklem et al. (2002)]{bar02} 
Barklem, P.~S., Stempels, H.~C., Allende Prieto, C., Kochukhov, O.~P.,
Piskunov, N., \& O'Mara, B.~J.\ 2002, \aap, 385, 951 
 
\bibitem[Beers et al. (1985)]{bee85} 
Beers, T.~C., Preston, G.~W., \& Shectman, S.~A.\ 1985, \aj, 90, 2089 

\bibitem[Bessell \& Norris (1984)]{ban84} 
Bessell, M.~S., \& Norris, J.\ 1984, \apj, 285, 622 

\bibitem[Bessell \& Brett (1988)]{bab88} 
Bessell, M. S., \& Brett, J. M. 1988, \pasp, 100, 1134

\bibitem[Bessell et al. (1991)]{bessel} 
Bessell, M.~S., Sutherland, R.~S., \& Ruan, K.\ 1991, \apjl, 383, L71 

\bibitem[Bessell, Castelli \& Plez (1998)]{bcp98} 
Bessell, M.~S., Castelli, F. \& Plez, B.\ 1998, \aap, 333, 231 

\bibitem[Boesgaard et al.(1999)]{Boesgaard1999} 
Boesgaard, A.~M., Deliyannis, C.~P., King, J.~R., Ryan, S.~G., Vogt,
S.~S., \& Beers, T.~C.\ 1999, \aj, 117, 1549 

\bibitem[Boesgaard et al. (1999)]{boe99} 
Boesgaard, A. M., King, J. R., Deliyannis, C. P., \& Vogt, S. S. 1999,
\aj, 117, 492 

\bibitem[Boesgaard \& Novicki(2006)]{Boesgaard2006} 
Boesgaard, A.~M., \& Novicki, M.~C.\ 2006, \apj, 641, 1122 

\bibitem[Bonifacio et al. (1990)]{bon89} 
Bonifacio, P., Castelli, F., Molaro, P. \ 
1990, in proceedings of ``Chemical and Dynamical Evolution of
Galaxies'', ed. F. Ferrini, J, Franco, F. Matteucci, p. 67
ETS Editrice, Pisa

\bibitem[Bonifacio \& Molaro (1997)]{BM97}
Bonifacio, P. \& Molaro, P. 1997, MNRAS, 285, 847

\bibitem[Bonifacio et al.(2002)]{B02} Bonifacio, P., et 
al.\ 2002, \aap, 390, 91 

\bibitem[Bonifacio et al. (2007)]{bon07} 
Bonifacio, P., Molaro, P., Sivarani, T. \AA. 2007, \aap, 462, 851
(Paper VII) 

\bibitem[Caffau et al.(2007)]{caffauS} 
Caffau, E., Faraggiana, R., Bonifacio, P., Ludwig, H.-G., \& Steffen,
M.\ 2007, \aap, 470, 699  

\bibitem[Caffau et al. (2007b)]{caffau07}
Caffau E., Ludwig, H.-G., Steffen, M., Ayres, T.R., Cayrel, R., Bonifacio, P.,
B. Freytag, B. Plez
\aap in preparation 
  
\bibitem[Carlsson et al. (1994)]{carlsson} 
Carlsson, M., Rutten, R.~J., Bruls, J.~H.~M.~J., \& Shchukina, N.~G.\
1994, \aap, 288, 860  

\bibitem[Carpenter(2001)]{carpenter} 
Carpenter, J.~M.\ 2001, \aj, 121, 2851 

\bibitem[Castelli \& Hubrig(2004)]{CH04} 
Castelli, F., \& Hubrig, S.\ 2004, \aap, 425, 263

\bibitem[{{Castelli} \& {Kurucz}(2003)}]{2003IAUS..210P.A20C}
{Castelli}, F. \& {Kurucz}, R.~L. 2003, in IAU Symposium, ed.
N.~{Piskunov}, W.~W. {Weiss}, \& D.~F. {Gray}, 20P

\bibitem[Cayrel (1988)]{cay88} 
Cayrel, R.\ 1988, IAU Symp.~132: The Impact of Very High S/N
Spectroscopy on Stellar Physics, 132, 345 

\bibitem[Cayrel \& Steffen (2000)]{CS00} 
Cayrel, R., \& Steffen, M.\ 2000, The Light Elements and their
Evolution, 198, 437 

\bibitem[Cayrel et al. (2004)]{cay04} 
Cayrel, R., Depagne, E., Spite, M. \AA. 2004, \aap, 416, 1117 (Paper V)

\bibitem[Cayrel et al. (2007)]{cay07} 
Cayrel, R., Steffen, M., Chand, H. \AA. 2007, \aap, 473, L37 

\bibitem[Charbonneau (1995)]{cha95} 
Charbonneau, P.\ 1995, \apjs, 101, 309 

\bibitem[Charbonnel \& Primas (2005)]{CP05} 
Charbonnel, C., \& Primas, F.\ 2005, \aap, 442, 961 

\bibitem[Cohen et al. (2004)]{coh04} 
Cohen, J. G., Christlieb, N., McWilliam, A., \AA. 2004, \apj, 612,
1107

\bibitem[Collet et al.(2007)]{collet} 
Collet, R., Asplund, M., \& Trampedach, R.\ 2007, \aap, 469, 687 

\bibitem[Deliyannis, Demarque, \& Kawaler (1990)]{deli} 
Deliyannis, C.~P., Demarque, P., \& Kawaler, S.~D.\ 1990, \apjs, 73, 21 

\bibitem[Edvardsson et al. (1993)]{edv93}
Edvardsson, B., Andersen, J., Gustafsson, B., Lambert, D.L., Nissen,
P.E., Tomkin, J. 1993, \aap, 275, 101  

\bibitem[Fran\c cois et al. (2003)]{fra03} 
Fran\c cois, P., Depagne, E., Hill, V., \AA. 2003, \aap, 403, 1105
(Paper III)
  
\bibitem[Fran{\c c}ois et al.(2004)]{fra04} 
Fran{\c c}ois, P., Matteucci, F., Cayrel, R., Spite, M., Spite, F., \&
Chiappini, C.\ 2004, \aap, 421, 613 

\bibitem[Frebel et al. (2005)]{fre05} 
Frebel, A., et al.\ 2005, \nat, 434, 871 

\bibitem[Freytag et al. (2002)]{fsd02} 
Freytag, B., Steffen, M., \& Dorch, B. 2002, Astronomische
Nachrichten, 323, 213 

\bibitem[Fuhrmann et al. (1993)]{fuh93}
Fuhrmann, K., Axer,  M., \& Gehren, T.\ 1993, \aap, 271, 451 


\bibitem[Gangrsky et al. (2006)]{HFSScII} 
Gangrsky, Y., et al.\ 2006, Hyperfine Interactions, 171, 209 

\bibitem[Gilmore et al.(1992)]{gilmore} 
Gilmore, G., Gustafsson, B., Edvardsson, B., \& Nissen, P.~E.\ 1992,
\nat, 357, 379  

\bibitem[Goldberg et al. (2002)]{gol02} 
Goldberg, D., Mazeh, T., Latham, D. W., \AA. 2002, \aj, 124, 1132

\bibitem[Grainge et al. (2003)]{gra03} 
Grainge, K., Carreira, P., Cleary, K., \AA. 2003, \mnras, 341, L23

\bibitem[Grevesse \& Sauval (2000)]{gas00} 
Grevesse, N., \& Sauval, A. 2000, Encyclopedia of Astronomy and
Astrophysics, Edited by Paul Murdin, article 1979. Bristol: Institute 
of Physics Publishing   

\bibitem[Gustafsson et al. (1975)]{gus75} 
Gustafsson, B., Bell, R.A., Eriksson, K., Nordlund \AA., 1975, \aap,
42, 407  

\bibitem[Gustafsson et al. (2003)]{gus03}
Gustafsson B.,  Edvardsson B., Eriksson K., Graae-J{\o}rgensen U.,
Mizuno-Wiedner, M., Plez, B., 2003, in Stellar Atmosphere Modeling, 
eds. I. Hubeny, D. Mihalas, K. Werner, ASP Conf. Series 288, 331. 

\bibitem[Harrington(1977)]{harrington} 
Harrington, R.~S.\ 1977, Revista Mexicana de Astronomia y Astrofisica,
vol.~ 3, 3, 139  

\bibitem[Israelian et al. (1998)]{isr98} 
Israelian, G., Garc{\'\i}a L\'opez, R. J., \& Rebolo, R. 1998, \apj,
507, 805 

\bibitem[Israelian et al. (2001)]{isr01} 
Israelian, G., Rebolo, R., Garc{\'\i}a L\'opez, R. J., Bonifacio,
P., Molaro, P., Basri, G., \& Shchukina, N. 2001, \apj, 551, 833

\bibitem[Khodykin et al.(2004)]{triples} 
Khodykin, S.~A., Zakharov, A.~I., \& Andersen, W.~L.\ 2004, \apj, 615,
506  

\bibitem[Korn et al.(2006)]{k2006} Korn, A.~J., Grundahl, F., 
Richard, O., Barklem, P.~S., Mashonkina, L., Collet, R., Piskunov, N., \& 
Gustafsson, B.\ 2006, \nat, 442, 657 

\bibitem[Korn et al.(2007)]{k2007} Korn, A.~J., Grundahl, F., 
Richard, O., Barklem, P.~S., Mashonkina, L., Collet, R., Gustafsson, B., \& 
Piskunov, N.\ 2007, ApJ, in press, arXiv:0709.0639 

\bibitem[{{Kurucz}(1993{\natexlab{a}})}]{1993KurCD..13.....K}
{Kurucz}, R. 1993{\natexlab{a}}, ATLAS9 
Stellar Atmosphere Programs and 2 km/s
grid.~Kurucz CD-ROM No.~13.~ Cambridge, Mass.: Smithsonian Astrophysical
Observatory, 1993., 13

\bibitem[{{Kurucz}(2005{\natexlab{a}})}]{2005MSAIS...8...14K}
{Kurucz}, R.~L. 2005{\natexlab{a}}, Memorie della Societ\`a
Astronomica Italiana Supplement, 8, 14

\bibitem[Kuo et al. (2004)]{kuo04} 
Kuo, C. L., Ade, P. A. R., Bock, J. J., \AA. 2004, \apj, 600, 32


\bibitem[Latham et al. (2002)]{lat02} 
Latham, D. W., Stefanik, R. P., Torres, G., \AA. 2002, \aj, 124, 1144

\bibitem[Lomb (1976)]{lom76} 
Lomb, N. R. 1976, \apss, 39, 447

\bibitem[Ludwig \& Steffen(2007)]{LS07} 
Ludwig, H.-G., \& Steffen, M.\ 2007, to appear in the Proceedings of
the ESO/Lisbon/Aveiro Workshop "Precision Spectroscopy in
Astrophysics", eds. L. Pasquini, M. Romaniello, N.C. Santos, and A.
Correia, arXiv:0704.1176  

\bibitem[McWilliam et al. (1995)]{mcw95} 
McWilliam, A., Preston, G.~W., Sneden, C., \& Searle, L.\ 1995, \aj, 109, 2757 

\bibitem[Mel{\' e}ndez \& Ram{\'{\i}}rez(2004)]{melendez} 
Mel{\' e}ndez, J., \& Ram{\'{\i}}rez, I.\ 2004, \apjl, 615, L33 

\bibitem[Molaro \& Castelli (1990)]{mac90} 
Molaro, P., \& Castelli, F. 1990, \aap, 228, 426
  
\bibitem[Molaro et al.(1997)]{mol97} 
Molaro, P., Bonifacio, P., Castelli, F., \& Pasquini, L.\ 1997, \aap,
319, 593  

\bibitem[Nissen (1989)]{nis89} 
Nissen, P. E. 1989, The Messenger, 58, 40

\bibitem[Nissen et al. (2002)]{nis02} 
Nissen, P. E., Primas, F., Asplund, M., \& Lambert, D. L. 2002, \aap,
390, 235

\bibitem[Norris et al. (1993)]{nor93} 
Norris, J. E., Peterson, R. C., Beers, T. C. 1993, \apj, 415, 797

\bibitem[Norris, Beers, \& Ryan (2000)]{nor00} 
Norris, J. E., Beers, T. C. \& Ryan, S. G. 2000, \apj, 540, 456

\bibitem[Pearson et al. (2003)]{pea03} 
Pearson, T. J., Mason, B. S., Readhead, A. C. S., \AA. 2003, \apj,
591, 556

\bibitem[Pickering(1996)]{pickering} 
Pickering, J.~C.\ 1996, \apjs, 107, 811 

\bibitem[Plez et al. (1992)]{ple92} 
Plez, B., Brett, J.M., Nordlund, \AA. 1992, \aap, 256, 551 
  
\bibitem[Preston et al. (1991)]{pre91} 
Preston, G. W., Shectman, S. A., Beers, T. C. 1991, \apjs, 76, 1001

\bibitem[Primas(2002)]{Primas02} 
Primas, F.\ 2002, \apss, 281, 195 

\bibitem[Richard et al.(2002)]{richard2002} Richard, O., Michaud, 
G., Richer, J., Turcotte, S., Turck-Chi{\`e}ze, S., \& VandenBerg, D.~A.\ 
2002, \apj, 568, 979 

\bibitem[Richard et al. (2005)]{richard} 
Richard, O., Michaud, G., Richer, J. 2005, \apj, 619, 538

\bibitem[Rebolo et al. (2004)]{reb04} 
Rebolo, R., Battye, R. A., Carreira, P., et al.
\AA. 2004, \mnras, 353, 747

\bibitem[Ryan et al.(1992)]{ryan92} 
Ryan, S.~G., Norris, J.~E., Bessell, M.~S., \& Deliyannis, C.\ 1992,
\apj, 388, 184  

\bibitem[Ryan, Norris, \& Beers(1999)]{ryan} 
Ryan, S.~G., Norris, J.~E., \& Beers, T.~C.\ 1999, \apj, 523, 654 

\bibitem[{{Sbordone}(2005)}]{2005MSAIS...8...61S}
{Sbordone}, L. 2005, Memorie della Societ\`a 
Astronomica Italiana Supplement, 8, 61

\bibitem[{{Sbordone} {et~al.}(2004){Sbordone}, {Bonifacio}, {Castelli}, \& {Kurucz}}]{2004MSAIS...5...93S}
{Sbordone}, L., {Bonifacio}, P., {Castelli}, F., \& {Kurucz}, R.~L.
2004, Memorie della Societ\`a Astronomica Italiana Supplement, 5, 93

\bibitem[Scargle (1982)]{sca82} 
Scargle, J. D. 1982, \apj, 263, 835

\bibitem[Schuster et al. (1993)]{sch93} 
Schuster, W. J., Parrao, L., \& Contreras Martinez, M. E. 1993, \aaps,
97, 951

\bibitem[Schuster et al. (1996)]{sch96} 
Schuster, W. J., Nissen, P. E., Parrao, L., Beers, T. C., \&
Overgaard, L. P. 1996, \aaps, 117, 317

\bibitem[Shchukina et al. (2005)]{Shc05} 
Shchukina, N.~G., Trujillo Bueno, J., \& Asplund, M.\ 2005, \apj, 618,
939  

\bibitem[Slettebak \& Brundage (1971)]{sle71} 
Slettebak, A., \& Brundage, R.~K.\ 1971, \aj, 76, 338 
 
\bibitem[Spergel et al. (2003)]{spe03} 
Spergel, D.~N., Bean, R., Dor\'e, O., \AA. 2003, \apjs, 148, 175 

\bibitem[Spergel et al. (2007)]{spe07} 
Spergel, D.~N., Verde, L., Peiris, H. V., \AA. 2007, \apjs, 170, 377

\bibitem[Spite \& Spite (1982a)]{sas82a} 
Spite, M.~\& Spite, F.\ 1982, \nat, 297, 483 

\bibitem[Spite \& Spite (1982b)]{sas82b} 
Spite, F., \& Spite, M.\ 1982, \aap, 115, 357 

\bibitem[van't Veer-Menneret \& M\'egessier (1996)]{vam96} 
van't Veer-Menneret, C., \& M\'egessier, C.\ 1996, \aap, 309, 879 

\bibitem[Wahlgren(2005)]{wahlgren} 
Wahlgren, G.~M.\ 2005, Memorie della Societ\`a Astronomica Italiana
Supplement, 8, 108  

\bibitem[Wedemeyer et al. (2003)]{wed03} 
Wedemeyer, S., Freytag, B., Steffen, M., Ludwig, H.-G., \& Holweger,
H. 2003, Astronomische Nachrichten, 324, 410

\bibitem[Wichmann et al. (2003)]{wic03} 
Wichmann, R., Schmitt, J. H. M. M., \& Hubrig, S. 2003, \aap, 400, 293

\end{thebibliography}
\end{document}